\documentclass[10pt,journal,twocolumn]{IEEEtran}

\usepackage{cite}
\usepackage{color}
\usepackage{amsmath}
\usepackage{amsfonts}
\usepackage{graphicx}
\usepackage{subfigure}
\usepackage{algorithm}
\usepackage{algorithmic}
\usepackage{cleveref}
\usepackage{epsfig}
  \usepackage{epstopdf}
\usepackage{setspace}
 \usepackage[noblocks]{authblk}
\usepackage{amssymb}

\newtheorem{theorem}{Theorem}

\begin{document}

\title{Joint Interleaver and Modulation Design For Multi-User SWIPT-NOMA}
\author{Yizhe Zhao,~\IEEEmembership{Student Member,~IEEE}, Jie Hu,~\IEEEmembership{Member,~IEEE}, Zhiguo Ding, ~\IEEEmembership{Senior Member,~IEEE}, and Kun~Yang,~\IEEEmembership{Senior Member,~IEEE}
\thanks{Yizhe Zhao, Jie Hu are with School of Information and Communication Engineering, University of Electronic Science and Technology of China, Chengdu, 611731, China, email: yzzhao@std.uestc.edu.cn, hujie@uestc.edu.cn.}
\thanks{Zhiguo Ding is with School of Electrical and Electronic Engineering, The University of Manchester, UK, e-mail: zhiguo.ding@manchester.ac.uk.}
\thanks{Kun Yang is with School of Computer Science and Electronic Engineering, University of Essex, Colchester, CO4 3SQ, U.K., and also with the School of Information and Communication Engineering, University of Electronic Science and Technology of China, Chengdu 611731, China, e-mail: kunyang@essex.ac.uk.}
\thanks{The financial support of the National Natural Science Foundation of China (NSFC), No. U1705263, and that of GF Innovative Research program, as well as that of the Sichuan Science and Technology Program, No. 2019YJ0194 are gratefully acknowledged. The work of Z. Ding was supported by the UK EPSRC under grant number EP/P009719/2 and by H2020-MSCA-RISE-2015 under grant number 690750.}
}

\maketitle

\begin{abstract}
Radio frequency (RF) signals can be relied upon for conventional wireless information transfer (WIT) and for challenging wireless power transfer (WPT), which triggers the significant research interest in the topic of simultaneous wireless information and power transfer (SWIPT). By further exploiting the advanced non-orthogonal-multiple-access (NOMA) technique, we are capable of improving the spectrum efficiency of the resource-limited SWIPT system. In our SWIPT system, a hybrid access point (H-AP) superimposes the modulated symbols destined to multiple WIT users by exploiting the power-domain NOMA, while WPT users are capable of harvesting the energy carried by the superposition symbols. In order to maximise the amount of energy transferred to the WPT users, we propose a joint design of the energy interleaver and the constellation rotation based modulator in the symbol-block level by constructively superimposing the symbols destined to the WIT users in the power domain. Furthermore, a transmit power allocation scheme is proposed to guarantee the symbol-error-ratio (SER) of all the WIT users. By considering the sensitivity of practical energy harvesters, the simulation results demonstrate that our scheme is capable of substantially increasing the WPT performance without any remarkable degradation of the WIT performance.
\end{abstract}


%

\section{Introduction}

\subsection{Background}

 In the upcoming era of 5G and Internet of Things (IoT), massive machine-type communications are enabled by the deployment of low-power IoT devices, which triggers more difficulties on the spectrum efficiency \cite{6464}, energy efficiency \cite{6565}, or security \cite{6060}. In order to accommodate the explosive growth of the machine-type tele-traffic, non-orthogonal-multiple-access (NOMA) \cite{777}\cite{1414} emerges as a critical technique for substantially increasing the spectrum efficiency. The NOMA aided transmitter differentiates the symbols destined to the different users by allocating various transmit power to these symbols superimposed in the same time-frequency resource block. The successive interference cancellation (SIC) technique \cite{2727} is adopted by the users for recovering their requested symbols from the superposition one. For a specific user,  the symbols requested by the other users are sequentially demodulated and removed from the superposition symbol until its own requested one is successfully recovered. In order to further increase the spectrum efficiency, the multiple-input-multiple-output (MIMO) technique \cite{2828} is combined with the NOMA by providing additional multiplexing gain in the spatial domain. By allowing the downlink and uplink transmissions in the same time-frequency resource block, the full-duplex technique may further double the spectrum efficiency of the NOMA \cite{2929}, if the self-interference can be appropriately mitigated.

Furthermore, in order to alleviate their energy shortage, radio frequency (RF) signal based wireless power transfer (WPT) can be relied upon for remotely charging these battery-powered IoT devices \cite{3636}. However, coordinating the conventional wireless information transfer (WIT) and the wireless power transfer (WPT) in the same RF band is a challenging task, which thus stimulates substantial research interest in the topic of simultaneous wireless information and power transfer (SWIPT) \cite{3737}. In a SWIPT systems, the simultaneous reception of both the information and energy is realised by invoking a signal splitter at the receiver, which is capable of splitting the received RF signal either in the time-domain \cite{2424}, or in the power-domain \cite{2525} or in the spatial domain \cite{3838}.  Furthermore, by exploiting the broadcast nature of the wireless channels, the transmitter is capable of simultaneously transferring RF signals to both the WIT users and the WPT users \cite{3939}. Specifically, the WIT users recover the requested information from the received RF signals, while the WPT users harvest energy from their received RF signals.

\subsection{Related Works}

Recently, some pioneering works have studied the multi-user SWIPT-NOMA system. For instance, Zheng \textit{et al.} \cite{888} considered a NOMA aided cooperative NOMA network, where a transmitter communicates with a pair of receivers via a single wireless powered relay. The optimal power allocation  scheme was found  for maximising the total downlink throughput. Moreover, Liu \textit{et al.} \cite{4141} proposed a novel protocol for the NOMA aided cooperative network, in which the users close to the transmitters operate as the wireless powered relays for the sake of forwarding the information to the distant users. Xu \textit{et al.} \cite{999} proposed a joint design of the transmit beamforming and receive power splitting in a similar NOMA aided cooperative network, which aims for maximising the achievable throughput of the so-called `strong users', while ensuring the minimum throughput requirements of other `weak users'. Diamantoulakis \textit{et al.} \cite{4040} studied a wireless-powered communication network, in which the uplink transmission of multi-user is supported by the NOMA.  Furthermore, Alsaba \textit{et al.} \cite{4242} studied a NOMA aided cooperative system, where the `strong user' adopts the full-duplex technique for simultaneously receiving the RF signal from the base station and for relaying information to the `weak user'. The sum-throughput was then maximised by invoking a two-step convex optimisation method. Mohammadali \textit{et al.} \cite{5151} studied a NOMA-SWIPT system having a battery powered access point and a pair of energy harvesting users. They investigated the information-energy trade-off, while studying the performance of cognitive radio aided NOMA. In their scheme,  the transmit power is allocated to the user having poor channel conditions in order to satisfy certain quality of service (QoS) requirements. Dai \textit{et al.} \cite{5252} investigated a millimeter-wave aided NOMA-SWIPT system on order to further enhance the spectrum efficiency. Moreover, Wei \textit{et al.} \cite{5353} proposed a transceiver design in a cooperative NOMA-SWIPT system, where the power allocation, the power splitting (PS) ratio and the receiver filter, as well as the transmit beamformer are jointly optimized  for maximising the SWIPT performance.

In the most of the existing works concerning the NOMA aided WIT and SWIPT systems, the WIT performance was always characterised by the classic Shannon-Hartley channel capacity, which inherently assumes infinite Gaussian-distributed continuous channel input. However, the attainable performance is always overestimated, since the channel input of any practical WIT or SWIPT system is normally finite and discrete \cite{6363}. As a result, we may attain more accurate performance when a practical modulator is considered. Cheng \textit{et al.} \cite{3333} studied the receive spatial modulation (RSM) aided SWIPT, where the optimal power allocation scheme between the information signal and the energy signal is optimised for achieving a balance between the information decoding and the energy harvesting requirements of the receiver. Mohjazi \textit{et al.} \cite{3434} analysed the end-to-end bit-error-ratio (BER) performance of the differential modulation in a SWIPT aided cooperation networks, where the wireless powered relay adopts the amplify-and-forward approach for relaying the information to the destination.  Unfortunately, few works really consider the modulation design of improving the SWIPT performance. There have been a number of works studying the modulation design in NOMA aided WIT systems \cite{4646}-\cite{4747}. Furthermore, Liu \textit{et al.} \cite{4545} proposed a beneficial amalgamation of the spatial modulation and the sparse code-division multiple-access (SCDMA), which is capable of supporting a very high user-load by exploiting the multiplexing gains provided by the multiple antennas and the non-orthogonality of the SCDMA.

In order to alleviate the impact of the cascaded errors imposed by bursty interference and deep fading, the symbols have to be shuffled and reordered by the interleaver before transmission. For instance, Li \textit{et al.} \cite{5555} studied a bit-interleaving scheme in space-time trellis coding in block-fading channels. Jun \textit{et al.} \cite{5656} proposed a serially concatenated continuous phase modulation by adopting symbol-level interleaving. On the other hand, constellation rotation was proposed for increasing the constrained capacity by rotating the constellations of different users with certain angles. In \cite{5757}, Neng \textit{et al.} studied constellation rotation of NOMA with SIC receiver, while the bit error rate was reduced and the capacity was enlarged compared with conventional NOMA. Lin \textit{et al.} \cite{5858} proposed an optimal inter-constellation rotation based on minimum distance criterion for uplink NOMA, while the robustness and fairness among users are both guaranteed.

\subsection{Contributions}

Introducing the WPT into the same RF band may make the precious spectrum even more congested. As a result, we have to exploit the high spectrum efficiency of the NOMA by satisfying both the WPT and WIT users' various requirements \cite{888}. However, superimposing the modulated RF signals destined to multiple users in the same time-frequency resource block may inevitably  degrade the actual energy carried by the resultant superposition signal, since these modulated signals having diverse phases may be destructively superimposed. Therefore, the resultant WPT performance is impaired. Recently, most existing works \cite{5151}-\cite{3333} for NOMA-SWIPT system mainly focus on the power allocation and the beamforming design for maximising the SWIPT performance, where the classic Shannon capacity is relied upon for characterising the upper-bound WIT performance. However, in a practical NOMA-SWIPT system with a practical modulation scheme having finite alphabet, this overestimated WIT performance  may result in unfair resource allocation among the WIT and WPT performance. Apart from the front-end transceiver and air-interface design, the potential of modulation design on improving the SWIPT performance has not been well understood. Unfortunately, few works studied the impact of modulation design in the NOMA-SWIPT system, except for our first attempt \cite{3535}. We proposed a constellation rotation modulation scheme in \cite{3535} for a three-user NOMA-SWIPT system, which includes a pair of WIT users and a single WPT user. The modulated signal destined to the WIT user pair is constructively superimposed in order to maximise its WPT performance. However, our design in \cite{3535} is only suitable for supporting a pair of WIT users, whose requested symbols are superimposed on a single sub-carrier (or a single resource block). In practice, we have to map symbols of different WIT users to the multiple sub-carriers available. Therefore, the energy interleaving is originally proposed for the sake of mitigating the destructive superposition of multiple symbols scheduled on each sub-carrier.

\begin{table*}
\setstretch{1.1}
\centering
\footnotesize
\caption{Definition of the key mathematical notations}
\label{tab:1}
\begin{tabular}{|c|l|c|l|}
\hline
\textbf{Symbols}     & \textbf{The meaning of these symbols} & \textbf{Symbols}     & \textbf{The meaning of these symbols}\\
\hline
$P_n$ & total transmit power on the $n$-th sub-carrier & $P_{n,k}$ & transmit power of $u^I_k$ on the $n$-th sub-carrier\\
\hline
$\mathbf{S}_{n,k}$ & the $n$-th symbol-block of $u^I_k$ & $s_{n,k,l}$ & the $l$-th symbol in $\mathbf{S}_{n,k}$\\
\hline
$A_{n,k,l}$ & normalised amplitude of $s_{n,k,l}$ & $\phi_{n,k,l}$ & phase of $s_{n,k,l}$\\
\hline
$\overline{\mathbf{S}}_{n,k}$ & the symbol-block of $u^I_k$ on the $n$-th sub-carrier & $\overline{s}_{n,k,l}$ & the $l$-th symbol in $\overline{\mathbf{S}}_{n,k}$\\
\hline
$\overline{A}_{n,k,l}$ & normalised amplitude of $\overline{s}_{n,k,l}$ & $\overline{\phi}_{n,k,l}$ & phase of $\overline{s}_{n,k,l}$\\
\hline
$\widehat{\mathbf{S}}_{n,k}$ & $\overline{\mathbf{S}}_{n,k}$ after constellation rotation & $\widehat{s}_{n,k,l}$ & the $l$-th symbol in $\widehat{\mathbf{S}}_{n,k}$\\
\hline
$\widehat{A}_{n,k,l}$ & normalised amplitude of $\widehat{s}_{n,k,l}$ & $\widehat{\phi}_{n,k,l}$ & phase of $\widehat{s}_{n,k,l}$\\
\hline
$\widehat{\mathbf{S}}_{n}$ & superimposed $\widehat{\mathbf{S}}_{n,k}$ of $K_I$ WIT users & $\widehat{s}_{n,l}$ & the $l$-th symbol in $\widehat{\mathbf{S}}_{n}$\\
\hline
$\widehat{x}_{n,l}$ & modulated signal of $\widehat{s}_{n,l}$ & $\widehat{E}_{n,l}$ & energy carried by $\widehat{x}_{n,l}$\\
\hline
$\mathbf{B}$ & energy interleaving matrix & $\boldsymbol{\Theta}$ & constellation rotation angle matrix\\
\hline
$\widehat{\xi}_{n,k,l}$ & constellation distortion of $\widehat{s}_{n,k,l}$ & $\eta_n$ & utility function on the $n$-th sub-carrier\\
\hline
\end{tabular}
\end{table*}
In order to improve the WPT performance of the NOMA aided SWIPT system, we aim for maximising the amount of  energy carried by the superposition signals by jointly designing the energy interleaver and the constellation rotation aided modulator of the transmitter in the symbol-block level, which may not remarkably degrade the WIT performance. To the best of our knowledge,  this is the first attempt to the interleaver and modulator design of the multi-user NOMA aided SWIPT system. Our novel contributions are summarised as below:
\begin{itemize}
\item A novel \textit{transceiver architecture} of the NOMA aided SWIPT system is proposed for simultaneously transferring energy to multiple WPT users and transferring information to multiple WIT users.
\item The transmitter design is obtained by jointly optimising the \textit{constellation rotation aided modulator} and by optimising the \textit{energy interleaver} for the sake of maximising the attainable WPT performance.
\item A feasible transmit \textit{power allocation scheme} among the WIT users is obtained without any remarkable degradation of the WIT performance in terms of symbol-error-ratio (SER).
\item The impact of the \textit{practical energy harvester's sensitivity} on the attainable WPT performance is also evaluated.
\end{itemize}

The rest of the paper is organized as follows: The transceiver architecture of the NOMA aided SWIPT system is introduced in Section II. The problem of the energy interleaver and the constellation rotator design is formulated in Section III, which is followed by the optimal design proposed in Section IV. The simulation results is then presented in Section V. The paper is finally concluded in Section VI.

\section{System Model and Transceiver Architecture}

Our NOMA aided SWIPT system consists of $K_I$ WIT users denoted as $\{u^I_1,\cdots,u^I_{K_I}\}$, $K_E$ WPT users denoted as $\{u^E_1,\cdots,u^E_{K_E}\}$ and a single hybrid access-point (H-AP), which is capable of simultaneously delivering information to the WIT users and transferring energy to the WPT users. All the users and the H-AP are equipped with a single antenna. The H-AP transmits the superposition signal to the WIT users by adopting the power-domain NOMA technique, while the WPT users harvest energy from this superposition signal by exploiting the broadcast nature of the wireless channels. The transmitter architecture of the H-AP, the receiver architecture of the WIT users and that of the WPT users are all portrayed in Fig.\ref{structure}. The definition of the key mathematical notations appearing in this paper are provided in TABLE \ref{tab:1}.

\subsection{Transmitter Architecture of The H-AP}

\begin{figure*}[t]
\setlength{\belowcaptionskip}{-1cm}
  \setlength{\abovecaptionskip}{0.2cm}
  \centering
  \includegraphics[width=0.8\linewidth]{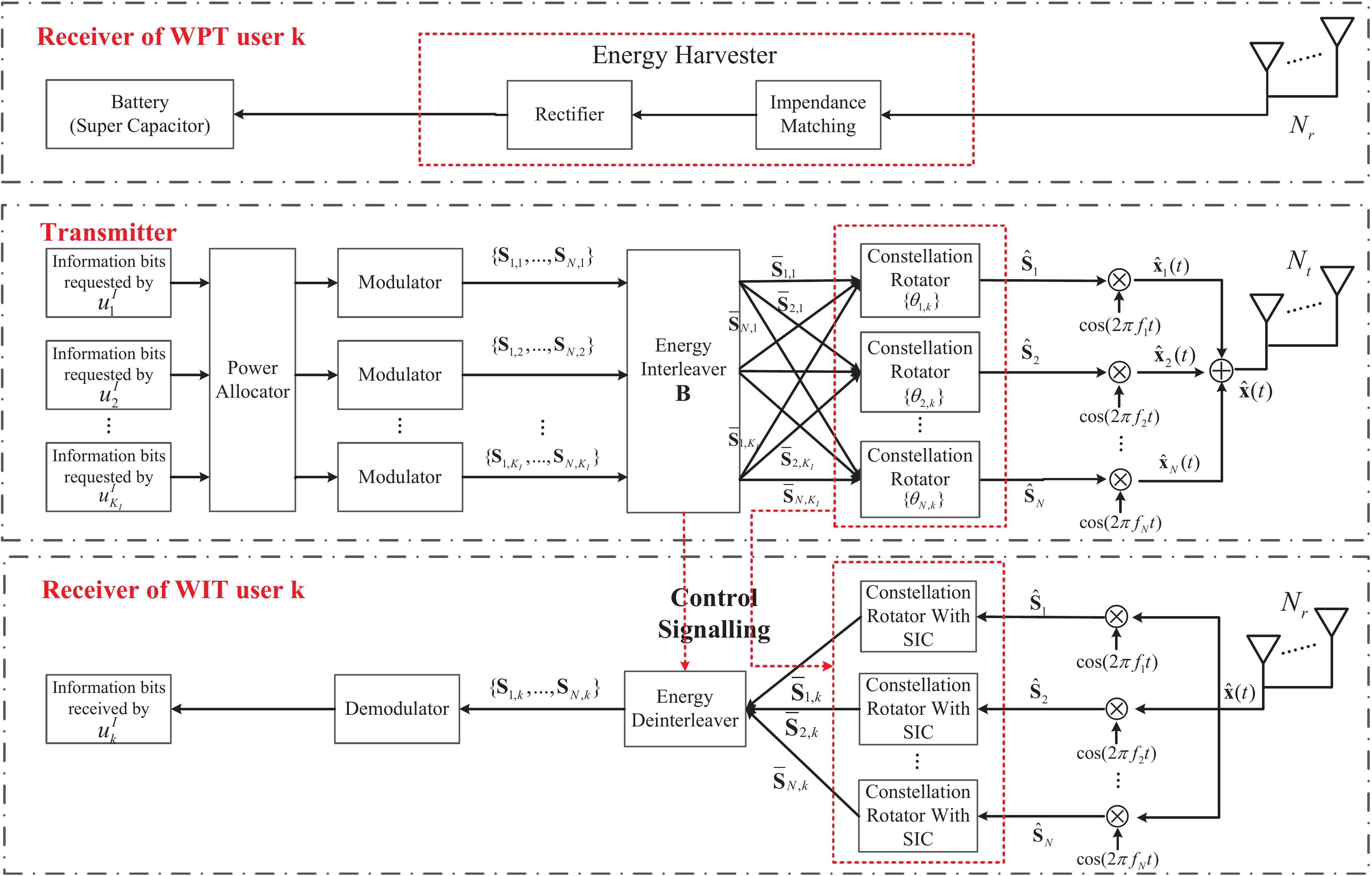}
  \caption{Transceiver architecture}\label{structure}
\end{figure*}

We illustrate the transmitter architecture of the H-AP in the midlle of Fig. \ref{structure}, where the H-AP transmits all requested symbols on $N$ different sub-carriers $\{\cos(2\pi f_n t)| n=1, \cdots, N\}$. The transmitter architecture includes the power allocator, the modulator, the energy interleaver and the constellation rotator.

\subsubsection{Power Allocator}

As illustrated in the middle of Fig. \ref{structure}, the symbols requested by the WIT user $\{u^I_k | k=1,\cdots, K_I\}$ are assigned allocated with different power $\{P_{n,k}|k=1,\cdots, K_I\}$ on the $n$-th sub-carrier, respectively. An appropriate power allocation scheme is capable of alleviating the mutual interference induced by the superposition symbol. More details will be provided in Section III-C.

\subsubsection{Modulator}
Since the WIT receivers require the knowledge of the energy interleaver and the constellation rotators for the sake of successful demodulation, this knowledge also needs to be transmitted to the receivers as the control overhead. In order to reduce the control overhead, both the energy interleaving and the constellation rotation operate in a symbol-block level. Each symbol-block consists of $L$ modulated  symbols. The transmitter generates $N*L$ modulated symbols for every WIT user, which are divided into $N$ symbol-blocks\footnote{The number of the symbol-blocks requested by a WIT user is equal to the number of sub-carriers.}. The symbol-blocks requested by WIT user $u^I_k$ are denoted as $\{\mathbf{S}_{n,k} | n=1,\cdots, N\}$, which is originally scheduled to be transmitted on the $n$-th carrier\footnote{This scheduling may be altered by the energy interleaver.}.

For the modulated symbols transmitted on the $n$-th sub-carrier to the WIT user $u^I_k$, the minimum Euclidean distance in the constellation is $2d_{n,k}$. Without loss of generality, M-QAM modulator is adopted as a case study in this manuscript, since it has been widely adopted in modern wireless communications, e.g. the Long Term Evolution (LTE) and 802.11 based system. Specifically, when the M-QAM is adopted, we have $d_{n,k}=\sqrt{\frac{3P_{n,k}}{M-1}}$ \cite{1111}. If the symbol-block $\mathbf{S}_{n,k}$ is transmitted to the WIT user $u^I_k$ on the $n$-th sub-carrier, the $l$-th symbol $s_{n,k,l}$ within it can be expressed as
\begin{align}\label{snkl}
s_{n,k,l}&=d_{n,k}\sqrt{(A^I_{n,k,l})^2+(A^Q_{n,k,l})^2}e^{j\phi_{n,k,l}}\notag\\
&=d_{n,k}A_{n,k,l}e^{j\phi_{n,k,l}},
\end{align}
for $l=1,\cdots, L$, where $\phi_{n,k,l}$ represents the phase, $A_{n,k,l}$ represents the amplitude normalised by $d_{n,k}$. Moreover,  $A^I_{n,k,l}$ and $A^Q_{n,k,l}$ are the normalised amplitudes of in-phase and quadrature in the constellation of M-QAM, respectively. Specifically, we have $A^I_{n,k,l}, A^Q_{n,k,l} \in \{1-\sqrt{M}, 3-\sqrt{M},\dots,\sqrt{M}-3, \sqrt{M}-1\}$.

\subsubsection{Energy Interleaver}

The symbol-blocks requested by all the WIT users flow into an energy interleaver for the following pair of purposes:
\begin{itemize}
	\item Converting the serial symbol-blocks requested by every WIT user into the parallel ones.
	\item Scheduling the transmissions of all the symbol-blocks on the $N$ sub-carriers. The scheduling scheme should avoid the superposition of the symbols having opposite phases, since it may degrade the actual energy carried by the resultant superposition symbols.
\end{itemize}
The energy interleaver can be represented by a three-dimensional binary tensor $\mathbf{B}=\{b_{k,m,n}\}$ having the size of $K_I\times N\times N$, where $b_{k,m,n}=1$ indicates that the symbol-block $\mathbf{S}_{m,k}$ is transmitted on the $n$-th sub-carrier, while $b_{k,m,n}=0$ indicates that $\mathbf{S}_{m,k}$ is not transmitted on the $n$-th sub-carrier. After the  energy interleaving, all the symbol-blocks are scheduled on the $N$ sub-carriers. The symbol-block requested by WIT user $u^I_k$ is denoted as $\overline{\mathbf{S}}_{n,k}$, which is transmitted on the $n$-th sub-carrier.

\subsubsection{Constellation Rotator}

After the energy interleaving,  the modulated symbol-blocks transmitted on every sub-carrier then flow into the respective constellation rotator, as portrayed in the middle of Fig. \ref{structure}.  The symbol-blocks requested by different WIT users are then rotated with some certain angles, respectively, in order to counteract the energy degradation induced by superimposing the symbols having opposite phases. After the constellation rotation, the symbol-block $\overline{\mathbf{S}}_{n,k}$ is then converted to $\widehat{\mathbf{S}}_{n,k}$. The constellation rotation angles are also represented by a matrix $\boldsymbol{\Theta}=\{\theta_{n,k}\}$ having the size of $N\times K_I$, in which $\theta_{n,k}$ indicates the constellation rotation angle of the symbol-block $\overline{\mathbf{S}}_{n,k}$.

Finally, all the  rotated symbol-blocks $\{\widehat{\mathbf{S}}_{n,k} | k=1,\cdots,K_I\}$ are superimposed together. The resultant superposition symbol-block is represented by $\widehat{\mathbf{S}}_{n}$. After being modulated onto the $n$-th sub-carrier, the corresponding signal $\widehat{x}_n(t)$ is then broadcast to all the WIT and WPT users.

\subsection{Receiver Architecture of WIT users}

The receiver architecture of the WIT users is illustrated in the bottom of Fig. \ref{structure}. At the receiver of WIT user $u^I_k$, the superposition symbol-blocks $\{\widehat{\mathbf{S}}_{1},\cdots,\widehat{\mathbf{S}}_N\}$ transmitted on $N$ sub-carriers are firstly recovered from the received analogue signal. Then, these symbol-blocks are further processed by the constellation rotation based SIC in order to extract the symbols requested by $u^I_k$, as presented in the bottom of Fig. \ref{structure}.

\subsubsection{Constellation Rotation Based SIC}

According to the classic SIC, the symbols having the higher power should be firstly demodulated. Without loss of generality, we assume $P_{n,1} < \cdots < P_{n,K_I}$ on each sub-carrier. Accordingly, the minimum Euclidean distances of the WIT users' constellations satisfy the inequality of $d_{n,1}<\cdots<d_{n,K_I}$. Then, the WIT user $u^I_k$ demodulates its requested symbol-block $\overline{\mathbf{S}}_{n,k}$ on the $n$-th sub-carrier by obeying the following steps:
\begin{itemize}
\item \textit{Step 1:} Initialize the label of the firstly demodulated WIT user as $i=K_I$;
\item \textit{Step 2:} Rotate the superposition symbol-block $\widehat{\mathbf{S}}_{n}$ with an angle of $-\theta_{n,i}$;
\item \textit{Step 3:} Recover the symbol-block $\overline{\mathbf{S}}_{n,i}$ according to the maximum-likelihood (ML) algorithm \cite{4949}, by regarding the symbol-blocks of all the other WIT users in $\widehat{\mathbf{S}}_{n}$ as the interference. If $i=k$, then $\overline{\mathbf{S}}_{n,i}$ is the symbol-block requested by $u^I_k$ on the $n$-th sub-carrier. Otherwise, remove $\overline{\mathbf{S}}_{n,i}$ from $\widehat{\mathbf{S}}_{n}$ and come back to Step 2 by setting $i=i-1$.
\end{itemize}

\subsubsection{Energy Deinterleaver}

After the constellation rotation based SIC, the recovered symbol-blocks $\overline{\mathbf{S}}_{n,k}$ of $u^I_k$ on all the sub-carriers are then processed by an energy deinterleaver for the following purposes:
\begin{itemize}	
\item Convert the parallel symbol-blocks to the serial ones.	
\item Sort the symbol-blocks to their original order according to $\mathbf{B}$ generated at the transmitter.
\end{itemize}

The sorted symbol-blocks after the energy deinterleaver are represented by $\{\mathbf{S}_{1,k},\cdots,\mathbf{S}_{N,k}\}$. After the demodulation, these symbols are finally decoded into information bits.

\subsection{Receiver Architecture of WPT users}
The receiver architecture of WPT users is illustrated in the top of Fig. \ref{structure}. The WPT users only harvest energy from the superposition signal destined to the WIT users by exploiting the broadcast nature of wireless channel. A typical energy harvester consists of an impendance matching circuit and a rectifier \cite{3737}. The impedance matching circuit guarantees an efficient energy transfer from the receive antennas to the electronic loads, while the rectifier converts the RF signals into the direct current (DC), whose energy can be finally stored in the battery or in the super capacitor.

The energy harvesters of the WPT users can only be activated, if the power of the received superposition signal is higher than a threshold $P_\mathrm{th}$, which is also regarded as the sensitivity of the energy harvester \cite{4848}. For simplicity, we assume an identical sensitivity for the energy harvesters of all the WPT. Specifically, the actual energy harvested by the WPT user $u^E_k$ within a symbol duration $T$ is $E^\mathrm{H}_k=P^\mathrm{H}_kT\cdot \mathbf{1}(P^\mathrm{H}_k\ge P_\mathrm{th})$, where $P^\mathrm{H}_k$ is the received power of the superposition signal, $\mathbf{1}(\cdot)$ is an indicator function, which is one only when the bracketed condition is satisfied, but zero, otherwise.

\subsection{Control signal}

\begin{figure}[t]
\centering
  \includegraphics[width=0.8\linewidth]{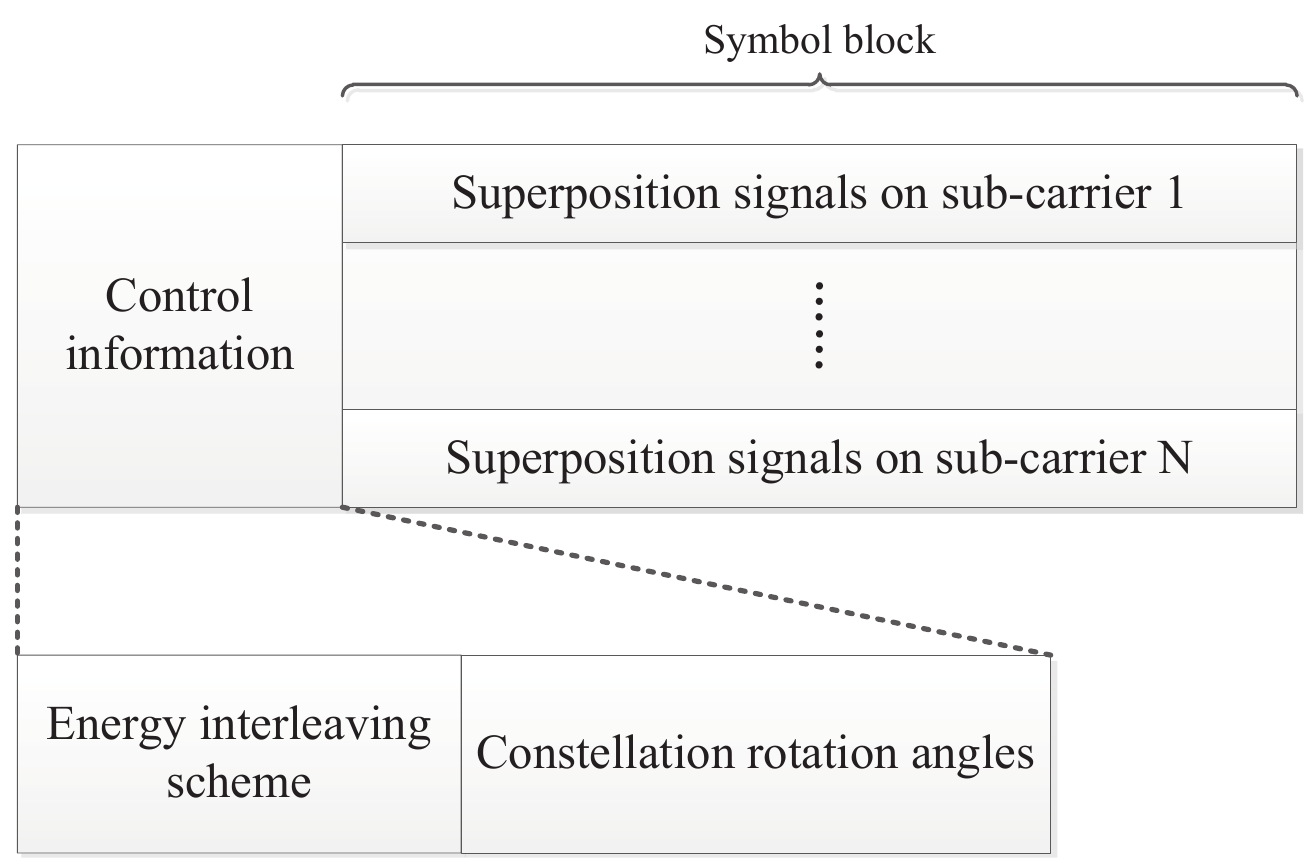}
  \caption{Control signal and data information frame}\label{frame}
\end{figure}

For the sake of symbol demodulation and of reordering the interleaved symbol blocks, the WIT users have to be aware of the energy interleaving schemes and constellation rotation angles  adopted by the transmitter. As a result, these control signalling overhead has to be transferred from the transmitter to the WIT users before actual data transmission, which is portrayed in Fig. \ref{frame}. These control signalling overhead can be compressed by adopting some appropriate encoding methods. For example, we have $N!$ symbol block allocation schemes in total for a specific WIT user, since the transmitter has to establish a one-on-one mapping between $N$ symbol block of this WIT user and $N$ sub-carriers. Therefore, the energy interleaving matrix $\mathbf{B}$ can be encoded into a binary sequence having a length of $K_I\lceil\log_2(N!)\rceil$, where $\lceil x\rceil$ indicates the minimum integer not lower than $x$. For the constellation rotation angles, we may divide the range of $[-\pi,\pi]$ into $D$ non-overlapping regions, namely $\bigcup_{k=1}^{D} [-\pi + \frac{2(k-1)\pi}{D}, -\pi + \frac{2k\pi}{D}]$. Since constellation rotation  angles are relative to one another, we may keep the constellation of an arbitrary WIT user unchanged while rotating the constellation of the other $(K_I-1)$ WIT users by certain angles. Therefore, all the constellation rotation  angles of the WIT users can be encoded as a binary sequence having a length of $N(K_I-1)\lceil\log_2(D)\rceil$.
\section{Specification of Key Functional Modules}

In this section, we will specify the key operations in the transmitter, which includes the energy interleaving, the constellation rotation and the power allocation.

\subsection{Energy Interleaving}

The energy interleaving is adopted at the transmitter in order to prevent the destructive superposition of the symbol-blocks requested by different WIT users on a single sub-carrier. Specifically, the WIT user $u^I_k$ requests $N$ symbol-blocks of $\{\mathbf{S}_{1,k}, \cdots, \mathbf{S}_{N,k}\}$ on $N$ sub-carriers, as portrayed in Fig. \ref{structure}. The transmitter has to carefully map these $N$ symbol-blocks onto the $N$ sub-carriers. Each symbol-block is only allowed to be scheduled on a single sub-carrier. The transmitter superimposes $K_I$ symbol-blocks requested by $K_I$ WIT users on every sub-carrier by adopting the power-domain NOMA. However, superimposing the symbols having various amplitudes and phases sometimes results in substantial degradation of the energy carried by the resultant superposition symbol, particularly when the symbols requested by the WIT users have opposite phases. As a result, we have to design an efficient interleaving scheme in order to appropriately map the symbol-blocks onto the sub-carriers and to avoid the degradation of the energy carried by the resultant superposition signal, which is regarded as the energy interleaving.

\begin{figure*}[t]
\setlength{\belowcaptionskip}{-1cm}
  \setlength{\abovecaptionskip}{0.2cm}
  \centering
  \includegraphics[width=0.8\linewidth]{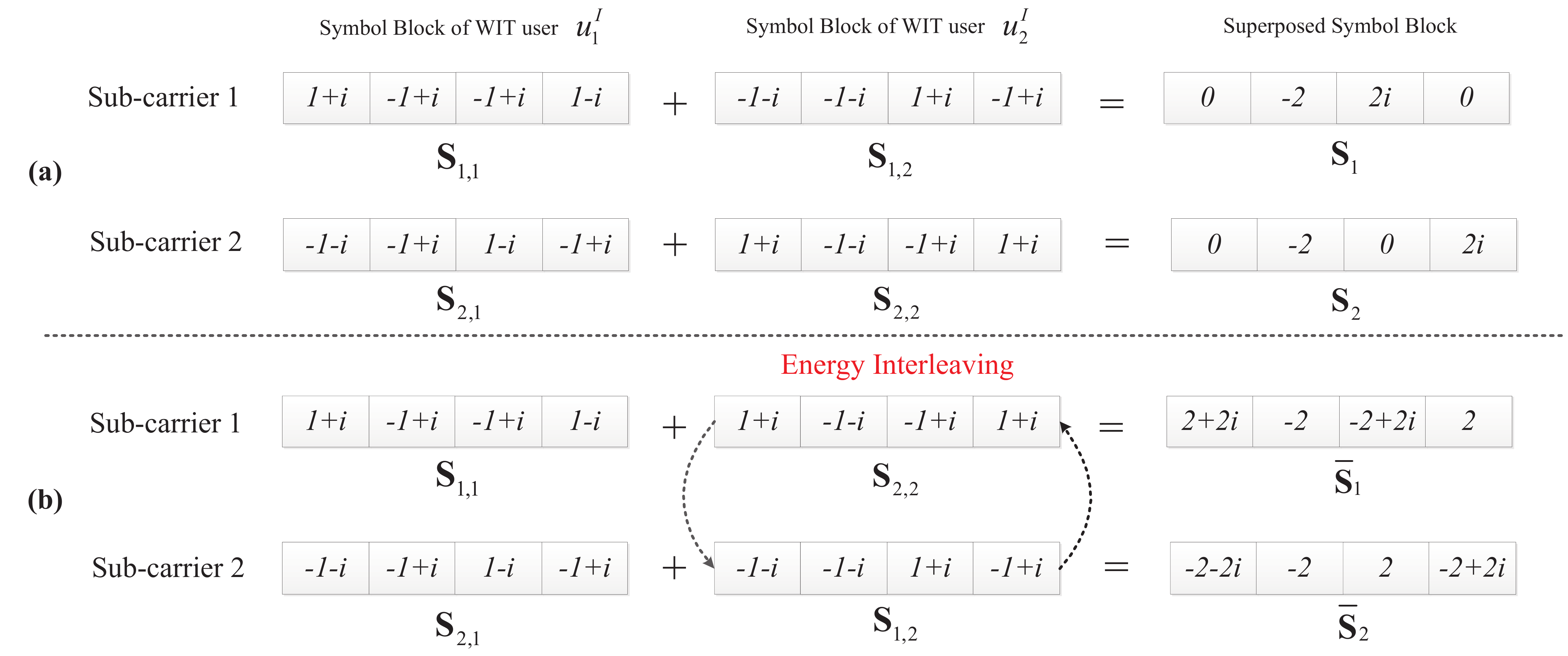}
  \caption{An example of energy interleaving having a pair of WIT users and a pair of subcarriers. Each WIT user requests for a pair of symbol-blocks, where the symbols are modulated by 4-QAM. The transmit power allocated to the symbols requested by the WIT user pair is identical.}\label{energyinterleaving}
\end{figure*}

Observe from Fig. \ref{energyinterleaving}(a) that when we do not adopt the energy interleaving, the energy carried by the superposition symbol-blocks $\mathbf{S}_1$ and $\mathbf{S}_2$ is substantially reduced. This is because some of the symbols in the blocks of $\mathbf{S}_{1,1}$ and $\mathbf{S}_{1,2}$ and some in $\mathbf{S}_{2,1}$ and $\mathbf{S}_{2,2}$ have opposite phases. By contrast, observe from Fig. \ref{energyinterleaving}(b) that when the energy interleaving is adopted, the transmitter may map the symbol-block $\mathbf{S}_{2,2}$ onto the first sub-carrier and map the symbol-block $\mathbf{S}_{1,2}$ onto the second sub-carrier. Therefore, the resultant superposition symbol-blocks $\overline{\mathbf{S}}_1$ and $\overline{\mathbf{S}}_2$ still carries considerable energy. As a result, the energy interleaving substantially increases the WPT performance of the superposition symbols.

According to the energy interleaving tensor $\mathbf{B}$, we have
\begin{align}
\overline{\mathbf{S}}_{n,k}=\sum\limits_{m=1}^Nb_{k,m,n}\mathbf{S}_{m,k}.
\end{align}
Furthermore, the $l$-th symbol $\overline{s}_{n,k,l}$  in the block $\overline{\mathbf{S}}_{n,k}$ can be expressed as
\begin{align}\label{eq:InterSymbol}
\overline{s}_{n,k,l}&=\sum\limits_{m=1}^Nb_{k,m,n}s_{m,k,l}=\sum\limits_{m=1}^Nb_{k,m,n}d_{n,k}A_{m,k,l}e^{j\phi_{m,k,l}}\notag\\
&=d_{n,k}\overline{A}_{n,k,l}e^{j\overline{\phi}_{n,k,l}}.
\end{align}

In \eqref{eq:InterSymbol}, $\overline{A}_{n,k,l}$ and $\overline{\phi}_{n,k,l}$ are the normalised amplitude and phase of the symbol $\overline{s}_{n,k,l}$, which are formulated as
\begin{align}\label{eq:Aphi}
\begin{cases}
\overline{\phi}_{n,k,l}=&\arctan\left(\frac{\sum\limits_{m=1}^NA_{m,k,l}b_{k,m,n}\sin(\phi_{m,k,l})}{\sum\limits_{m=1}^NA_{m,k,l}b_{k,m,n}\cos(\phi_{m,k,l})}\right),\\
\overline{A}_{n,k,l}=&\left[\left(\sum\limits_{m=1}^NA_{m,k,l}b_{k,m,n}\cos(\phi_{m,k,l})\right)^2\right.\\
&\left.+\left(\sum\limits_{m=1}^NA_{m,k,l}b_{k,m,n}\sin(\phi_{m,k,l})\right)^2\right]^\frac{1}{2}.
\end{cases}
\end{align}

\subsection{Constellation Rotation}

Constellation rotation aims for avoiding the destructive superposition among the symbol-blocks transmitted on a single sub-carrier by rotating the symbol blocks requested by different WIT users with different angles. As a result, the WPT performance of the resultant superposition symbol-block can be substantially improved.

\begin{figure*}[t]
\setlength{\belowcaptionskip}{-1cm}
  \setlength{\abovecaptionskip}{0.2cm}
  \centering
  \includegraphics[width=0.8\linewidth]{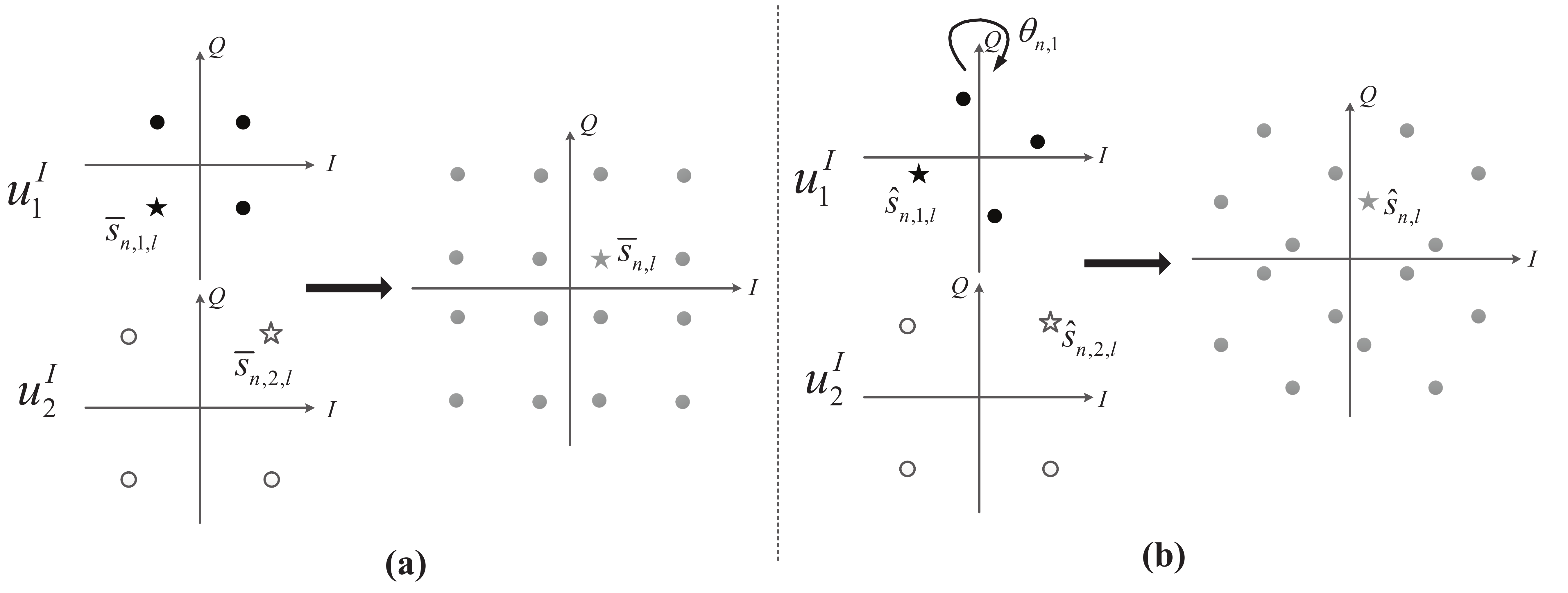}
  \caption{Examples of the symbol superposition between a pair of WIT users: (a) Without constellation rotation; (b) With constellation rotation. The modulated symbols are obtained by invoking 4-QAM. The symbols requested by the WIT user $u_1^I$ have a lower transmit power than those requested by the WIT user $u_2^I$. The symbol requested by the WIT user $u_1^I$ is denoted as the filled star, while the symbol requested by $u_2^I$ is denoted as the unfilled star. The resultant superposition symbol is denoted as the grey star. }\label{constellation rotation}
\end{figure*}

The symbol superposition between a pair of WIT users with and without the constellation rotation has been exemplified in Fig. \ref{constellation rotation}. Specifically, observe from Fig. \ref{constellation rotation}(a) that the actual power carried by the superposition symbol $\overline{s}_{n,l} = \overline{s}_{n,1,l} + \overline{s}_{n,2,l}$ is lower than the sum power of the individual symbols $\overline{s}_{n,1,l}$ and $\overline{s}_{n,2,l}$. The opposite phases of $\overline{s}_{n,1,l}$ and $\overline{s}_{n,2,l}$ substantially impair the attainable WPT performance. By contrast, if we rotate $u^I_1$'s constellation with a certain angle $\theta_{n,1}$, while keeping the constellation of the WIT user $u^I_2$ unchanged, we may observe from Fig. \ref{constellation rotation} that the resultant superposition symbol $\widehat{s}_{n,l} = \widehat{s}_{n,1,l} + \widehat{s}_{n,2,l}$ after the constellation rotation carries a higher power than its counterpart $\overline{s}_{n,l}$.

After the constellation rotation, the symbol-block $\overline{\mathbf{S}}_{n,k}$ is converted into
\begin{align}
\widehat{\mathbf{S}}_{n,k}=\overline{\mathbf{S}}_{n,k}e^{j\theta_{n,k}},
\end{align}
where $\theta_{n,k}$ represents the rotation angle for all the symbols in the symbol-block $\overline{\mathbf{S}}_{n,k}$. The $l$-th symbol $\widehat{s}_{n,k,l}$ in $\widehat{\mathbf{S}}_{n,k}$ is thus formulated as
\begin{align}
\widehat{s}_{n,k,l}=\overline{s}_{n,k,l}e^{j\theta_{n,k}}=d_{n,k}\overline{A}_{n,k,l}e^{j(\overline{\phi}_{n,k,l}+\theta_{n,k})},
\end{align}

Then, after the superposition of the $K_I$ WIT users' symbol-blocks, the resultant superposition symbol-block $\widehat{\mathbf{S}}_n$ transmitted on the $n$-th sub-carrier is formulated as
\begin{align}
\widehat{\mathbf{S}}_n=\sum\limits_{k=1}^{K_I}\widehat{\mathbf{S}}_{n,k}.
\end{align}
Moreover, the $l$-th superposition symbol $\widehat{s}_{n,l}$ in $\overline{\mathbf{S}}_n$ is derived as
\begin{align}
\widehat{s}_{n,l}=\sum\limits_{k=1}^{K_I}\widehat{s}_{n,k,l}=\sum\limits_{k=1}^{K_I}d_{n,k}\widehat{A}_{n,k,l}e^{j\widehat{\phi}_{n,k,l}},
\end{align}
where  we have $\widehat{A}_{n,k,l}=\overline{A}_{n,k,l}$ and $\widehat{\phi}_{n,k,l}=\overline{\phi}_{n,k,l}+\theta_{n,k}$.

Finally, the modulated RF signal $\widehat{x}_{n,l}(t)$ corresponding to the superposition symbol $\widehat{s}_{n,l}$ can be formulated as
\begin{align}\label{eq:3}
\widehat{x}_{n,l}(t)&=\text{Re}\left[\widehat{s}_{n,l}e^{j2\pi f_nt}\right]\notag\\
&=\sum\limits_{k=1}^{K_I}d_{n,k}\widehat{A}_{n,k,l}\cos(2\pi f_n t+\widehat{\phi}_{n,k,l}),\notag\\
&\ \ \ \ \ \ \ \ \ \ \ \ \ \ \ \ \ \ \ \ \ \ \ (l-1)T \le t \le lT.
\end{align}

\subsection{Power Allocation for WIT Assurance}

\begin{figure}[t]
  \centering
  \includegraphics[width=1\linewidth]{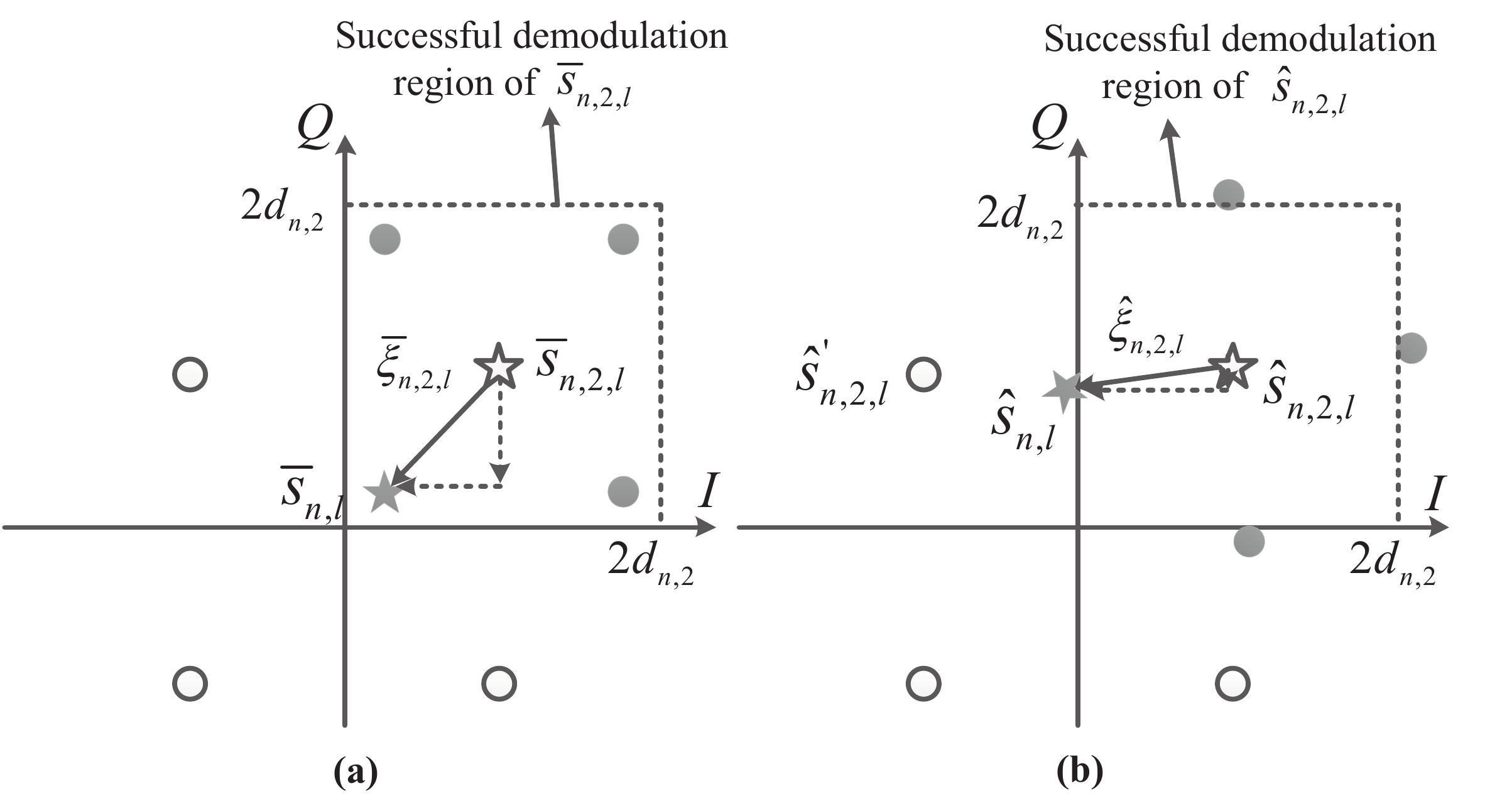}
  \setlength{\belowcaptionskip}{-12pt}
  \caption{The successful demodulation region of the WIT user $u_2^I$'s symbol: (a) without constellation rotation; (b) with constellation rotation. The assumptions are in line with those made in Fig.\ref{constellation rotation}.}\label{TAC}
\end{figure}

Although the attainable WPT performance can be increased by the constellation rotation, it may incur the adverse symbol offset, which may degrade the associated WIT performance. According to the SIC, when demodulating the symbol $\overline{s}_{n,2,l}$ (or $\widehat{s}_{n,2,l}$), its counterpart $\overline{s}_{n,1,l}$ (or $\widehat{s}_{n,1,l}$) is regarded as interference, as exemplified in Fig. \ref{TAC}. Observe from Fig. \ref{TAC}(a) that without the constellation rotation, the superposition symbol $\overline{s}_{n,l}$ is still within the successful demodulation region of $\overline{s}_{n,2,l}$. By contrast, in the case of the constellation rotation, the superposition symbol $\widehat{s}_{n,l}$ is drifted beyond the successful demodulation region of $\widehat{s}_{n,2,l}$, as presented in Fig. \ref{TAC}(b), which substantially degrades the demodulation performance of the symbol $\widehat{s}_{n,2,l}$. Since we adopt the SIC for sequentially demodulating the symbols requested by different WIT users, we may suffer from severe error propagation when we demodulate the symbols of the rest of WIT users.

According to the SIC , after demodulating the symbol-blocks $\{\widehat{\mathbf{S}}_{n,k+1},\cdots,\widehat{\mathbf{S}}_{n,K_I}\}$ requested by other WIT users and remove them from the superposition symbol, the WIT user $u^I_k$ then rotates the residual superposition symbol with an angle of $-\theta_{n,k}$ for demodulating its own requested symbol-block $\widehat{\mathbf{S}}_{n,k}$, according to the SIC specified in Section II-B-1. As a result, the distortion induced by the $l$-th symbols belonging to the blocks $\{\widehat{\mathbf{S}}_{n,1},\cdots,\widehat{\mathbf{S}}_{n,k-1}\}$ on the $l$-th symbol belonging to the block $\widehat{\mathbf{S}}_{n,k}$ can be formulated as
\begin{align}\label{xi}
\widehat{\xi}_{n,k,l}&=\sum\limits_{i=1}^{k-1}\widehat{s}_{n,i,l}e^{-j\theta_{n,k}}\notag\\
&=\sum\limits_{i=1}^{k-1}d_{n,i}\widehat{A}_{n,i,l}\cos(\widehat{\phi}_{n,i,l}-\theta_{n,k})\notag\\
&\ \ +j\sum\limits_{i=1}^{k-1}d_{n,i}\widehat{A}_{n,i,l}\sin(\widehat{\phi}_{n,i,l}-\theta_{n,k})\notag\\
&\triangleq \ \widehat{\xi}_{n,k,l}^I+j\widehat{\xi}_{n,k,l}^Q,
\end{align}
where $\widehat{\xi}_{n,k,l}^I$ represents the in-phase component of the distortion vector $\widehat{\xi}_{n,k,l}$ and  $\widehat{\xi}_{n,k,l}^Q$ represents the corresponding component in the quadrature phase. In order to reduce the SER of the WIT user $u^I_k$, the distortion induced by the symbol superposition should be carefully controlled. Therefore, the residual superposition symbol should be still within the successful demodulation region of the symbol requested by the WIT user $u^I_k$, which can be represented by a square having the side length of $2d_{n,k}$, as exemplified in Fig.\ref{TAC}. The constraints on the symbol distortion is then formulated as
 \begin{align}\label{eq:7}
|\widehat{\xi}_{n,k,l}^I|,|\widehat{\xi}_{n,k,l}^Q|\le d_{n,k},
 \end{align}
for $l=1,\cdots,L$, $k=2,\cdots,K_I$, $n=1,\cdots,N$, where $|x|$ indicates the absolute value of the complex number $x$.

In order to satisfy the symbol distortion constraint of \eqref{eq:7}, we opt to control the transmit power $\{P_{n,1}, \cdots, P_{n,K_I}\}$ allocated to the symbols requested by all the $K_I$ WIT users. As a result, the constellation rotation angles may gain the highest degree of freedom for the sake of maximising the energy carried by the superposition symbol. The following theorem thus provides a feasible power allocation scheme:
\begin{theorem}
In order to satisfy the symbol distortion constraint of \eqref{eq:7},  the transmit power $\{P_{n,1},\cdots,P_{n,K_I}\}$ of the symbols requested by all the WIT users should satisfy the following inequality:
\begin{align}\label{eq:12}
\sqrt{P_{n,k}}&\ge A_{\max}\sum_{i=1}^{k-1}\sqrt{P_{n,i}},
\end{align}
for $k=2,\cdots,K_I$, where $A_{\max}$ represents the normalised maximum amplitude, when the M-QAM is adopted for generating the modulated symbols.
\end{theorem}

\begin{IEEEproof}
Please refer to Appendix A for detailed proof.
\end{IEEEproof}

In order to guarantee the fairness among users but to satisfy the symbol distortion constraint \eqref{eq:7}, we take the equality of \eqref{eq:12}. Therefore, the transmit power of the symbols requested by the WIT user $u^I_k$ can be expressed as
\begin{align}\label{Pkconstraint2}
\sqrt{P_{n,k}}=A_{\max}\sum\limits_{i=1}^{k-1}\sqrt{P_{n,i}}, \ k=2,\cdots,K_I.
\end{align}

Given the total transmit power $P_n=\sum_{k=1}^{K_I}P_{n,k}$ on the $n$-th sub-carrier, the transmit power of the symbols requested by the WIT users can be thus formulated as
\begin{align}\label{nofairpowerallocation}
	P_{n,k} = \begin{cases}
    \left(1+\sum\limits_{k=2}^{K_I}\left(\frac{A_{\max}-A_{\max}^k}{1-A_{\max}}\right)^2\right)^{-1}P_n, k=1\\
	A_{\max}\sum\limits_{i=1}^{k-1}\sqrt{P_{n,i}}, k=2,\cdots, K_I
	\end{cases}.
\end{align}

\section{Optimal Design of Energy Interleaver and Constellation Rotator}

\subsection{Problem Formulation}

The actual energy $\widehat{E}_{n,l}$ carried by the $l$-th superposition signal $\widehat{x}_{n,l}(t)$ transmitted to all the WIT users on the $n$-th sub-carrier can be expressed as \eqref{eq:Enl}.
\begin{figure*}[htb]
\begin{align}\label{eq:Enl}
\widehat{E}_{n,l}=Tf_n\displaystyle{\int_{-\frac{1}{2f_n}}^{\frac{1}{2f_n}}\left(\widehat{x}_{n,l}(t)\right)^2dt}=\frac{T}{2}\left(\sum\limits_{k_1=1}^{K_I}\sum\limits_{k_2=1}^{K_I}d_{n,k_1}d_{n,k_2}\widehat{A}_{n,k_1,l}\widehat{A}_{n,k_2,l}\cos(\widehat{\phi}_{n,k_1,l}-\widehat{\phi}_{n,k_2,l})\right)
\end{align}
\end{figure*}

Given the specific power allocation scheme \eqref{nofairpowerallocation} among the symbols requested by different WIT users, the joint design of the energy interleaver $\mathbf{B}$ and the constellation rotator $\boldsymbol{\Theta}$ for maximising the energy carried by the superposition symbols on the multi-carriers can be formulated as
 \begin{align}
&\text{(P1): } \max_{\boldsymbol{\Theta},\mathbf{B}} \ \ \ \ \ \ \ \ \ \ \ \ \ \ \ \ \widehat{E}=\sum\limits_{n=1}^N\sum\limits_{l=1}^L\widehat{E}_{n,l},\label{P1}\\
&\text{s. t.}\ \ \ -\pi\le\theta_{n,k}<\pi,\ 1\le k\le K_I, \ \ 1\le n\le N,\tag{16a}\label{P1-A}\\
 &\ \ \ \ \ \ \ b_{k,m,n}=0\ or\ 1,\ \ 1\le k\le K_I, \ \ 1\le n\le N, \notag\\
 &\ \ \ \ \ \ \ \ \ \ \ \ \ \ \ \ \ \ \ \ \ \ \ \ \ \ \ \ \ \ \ \ \ \ \ \ \ \ \ \ \ \ \ \ \ \ 1\le m\le N, \tag{16b}\label{P1-B}\\
 &\ \ \ \ \ \ \sum_{m=1}^Nb_{k,m,n}=1,\ \ \ 1\le k\le K_I, \ \ 1\le n\le N,\tag{16c}\label{P1-C}\\
 &\ \ \ \ \ \ \sum_{n=1}^Nb_{k,m,n}=1,\ \ \ 1\le k\le K_I, \ \ 1\le m\le N,\tag{16d}\label{P1-D}
\end{align}
where \eqref{P1-A} and \eqref{P1-B} constraint the optional range of $\mathbf{B}$ and $\boldsymbol{\Theta}$, \eqref{P1-C} and \eqref{P1-D} indicate that for a single WIT user $u^I_k$, only a single symbol-block is allowed to be transmitted on a single sub-carrier.

The optimisation problem (P1) is a mixture of integer and real number programming. It's impossible to obtain a closed-form solution. This problem can be addressed by sequentially solving sub-problem of the constellation rotator design and that of the energy interleaver design. Note that our design actually maximises the total energy carried by the multi-carrier signal $\widehat{\mathbf{x}}(t)$ before it goes to the radio front, as illustrated in Fig.\ref{structure}. Therefore, our design may naturally adapt to any arbitrary number of antennas.

\subsection{Constellation Rotator Design}

\subsubsection{Utility Function}
Given a specific energy interleaver $\mathbf{B}$, the original optimisation problem (P1) can be simplified as
\begin{align}
\text{(P2): }\max_{\boldsymbol{\Theta}} \ \ \ \ \widehat{E}=\sum\limits_{n=1}^N\sum\limits_{l=1}^L\widehat{E}_{n,l},\ \ \ \ \text{s. t. }\ \ \ \ \eqref{P1-A}.\label{P2}
\end{align}
As portrayed in Fig.\ref{structure}, every sub-carrier has a constellation rotator. Therefore, the constellation rotator design on a specific sub-carrier is independent with others. As a result, the optimisation problem (P2) can be further decomposed into $N$ sub-problems in order to obtain the optimal constellation rotation angles for every WIT user on the $N$ sub-carriers, respectively. Specifically, the optimal constellation rotator design on the $n$-th sub-carrier can be formulated as
\begin{align}
\text{(P3): }\max_{\{\theta_{n,1},\cdots,\theta_{n,K_I}\}} \ \ \widehat{E}_n=\sum\limits_{l=1}^L\widehat{E}_{n,l},\ \ \text{s. t. }\ \ \ \ \eqref{P1-A}.\label{P3}
\end{align}
Given the symbol-blocks $\{\overline{\mathbf{S}}_{n,1},...,\overline{\mathbf{S}}_{n,K_I}\}$ transmitted on the $n$-th sub-carrier, the utility function of this sub-carrier can be defined as $\eta_n(\overline{\mathbf{S}}_{n,1},...,\overline{\mathbf{S}}_{n,K_I}) = \widehat{E}_{n,\max}$, where $\widehat{E}_{n,\max}$ is the resultant maximum by solving the problem (P3).

Observe from (P3) and \eqref{eq:Enl} that the non-convex cosine function exists in the objective of \eqref{P3}. As a result, the optimisation problem (P3) is not convex and we cannot solve it by exploiting the convex optimisation. However, we may obtain its sub-optimal solution by exploiting the method of alternating optimisation \cite{5050}.

\subsubsection{Iterative Algorithm}

The iterative algorithm for solving (P3) has the following steps:
\begin{itemize}
    \item \textit{Step 1}: Randomly initialise the constellation rotation angles $\{\theta_{n,1},\cdots,\theta_{n,K_I}\}$ according to the constraint \eqref{P1-A}.
    \item \textit{Step 2}: Optimise the constellation rotation angle $\theta_{n,k}$ for the symbols requested by the WIT user $u^I_k$ for $k=1,\cdots,K_I$ by solving (P3), when the other angles $\{\theta_{n,1},\cdots,\theta_{n,K_I}\}/\{\theta_{n,k}\}$ are regarded as constants. According to the trigonometry, the optimal rotation angle $\theta_{n,k}$ is derived as	
	\begin{align}\label{eq:thetank0}
	\theta_{n,k}=-\arctan\left(\frac{\sum\limits_{l=1}^L\sum\limits_{\substack{i=1,i\neq k}}^{K_I}\alpha_{n,i,k,l}\sin(\beta_{n,i,k,l})}{\sum\limits_{l=1}^L\sum\limits_{\substack{i=1,i\neq k}}^{K_I}\alpha_{n,i,k,l}\cos(\beta_{n,i,k,l})}\right),
	\end{align}
where $\alpha_{n,i,k,l}=d_{n,k}d_{n,i}\overline{A}_{n,k,l}\overline{A}_{n,i,l}$, and $\beta_{n,i,k,l}=\overline{\phi}_{n,k,l}-\overline{\phi}_{n,i,l}-\theta_{n,i}$.
	\item \textit{Step 3}: Sequentially optimise the rotation angles for all the WIT users by using the same method of \textit{Step 2}.
	\item \textit{Step 4}: Repeat Steps 3 and 4 until the constellation rotation angles converge to their optimums $\{\theta_{n,1}^*,\cdots,\theta_{n,K_I}^*\}$.
\end{itemize}
The pseudo code of the iterative algorithm is then provided in Algorithm \ref{alg1}.

\begin{algorithm}[t]
\setlength{\belowcaptionskip}{-1cm}
\setstretch{1.0}
\caption{Iterative algorithm for obtaining constellation rotation angles.}
\label{alg1}
\small
\textbf{INPUT:}
User number $K_I$;\
Symbol blocks $\overline{\mathbf{S}}_{n,1},\cdots,\overline{\mathbf{S}}_{n,K_I}$;\
Convergence accuracy $\Lambda$;\\
\textbf{OUTPUT:}
Constellation rotation angles $\theta_{n,1}^*,\cdots,\theta_{n,K_I}^*$.
\begin{algorithmic}[1]
\STATE \textbf{Initialise} Iteration counter $q\leftarrow 1$; $\theta_{n,1}^{(0)}=\cdots=\theta_{n,K_I}^{(0)}\leftarrow 0$; Randomly choose $\theta_{n,1}^{(1)},\cdots,\theta_{n,K_I}^{(1)}$ within the range of $[-\pi,\pi)$; Iteration user $k\leftarrow 1$;
\WHILE {Not all $|\theta_{n,k}^{(q)}-\theta_{n,k}^{(q-1)}|<\Lambda$}
\WHILE {$k\le K_I$}
\STATE Obtain $\theta_{n,k}$ according to \eqref{eq:thetank0}, $\theta_{n,k}^{(q+1)}\leftarrow \theta_{n,k}$;
\STATE $k\leftarrow k+1$;
\ENDWHILE
\STATE $k\leftarrow 1$, $q\leftarrow q+1$;
\ENDWHILE
\STATE Obtain $\theta_{n,k}^*\leftarrow \theta_{n,k}^{(q)}$ for $k=1,\cdots,K_I$;
\end{algorithmic}
\end{algorithm}

In Step 2 of every iteration, we may obtain the optimal solution of $\theta_{n,k}$, although the problem is non-convex. We let $E_n(\theta_{n,k}^{(q)})$ represent the energy carried by the superposition symbol on the $n$-th sub-carrier, when $\theta_{n,k}^{(q)}$ is updated at the $q$-th iteration according to Algorithm \ref{alg1}. The following inequalities can be readily obtained:
\begin{align}
&E_n(\theta_{n,k}^{(q)})\le E_n(\theta_{n,k+1}^{(q)})\le \cdots\le E_n(\theta_{n,K_I}^{(q)})\notag\\
&\le E_n(\theta_{n,1}^{(q+1)})\le\cdots\le E_n(\theta_{n,k}^{(q+1)}).
\end{align}
This is because in every round of optimisation, the objective of (P3) is a non-decreasing function. Furthermore, according to \eqref{eq:Enl}, we also have
\begin{align}
E_n(\theta_{n,k}^{(q)})\le& \sum\limits_{l=1}^{L}\frac{T}{2}\left(\sum\limits_{k_1=1}^{K_I}\sum\limits_{k_2=1}^{K_I}d_{n,k_1}d_{n,k_2}\overline{A}_{n,k_1,l}\overline{A}_{n,k_2,l}\right),\notag\\
&\forall 1\le k\le K_I, q=1,2,\cdots,
\end{align}
which indicates that $\{E_n(\theta_{n,k}^{(q)})\}$ is an increasing function with respect to the iteration round $q$, while $\{E_n(\theta_{n,k}^{(q)})\}$ is also upper-bounded. Therefore, Algorithm \ref{alg1} finally converges as the iteration round $q$ increases. When the algorithm converges, it returns us the sub-optimal solution.

\subsection{Energy Interleaver Design}

The utility of a specific symbol-blocks' transmission arrangement $\{\overline{\mathbf{S}}_{n,1},...,\overline{\mathbf{S}}_{n,K_I}\}$ on the $n$-th sub-carrier can be characterised by the function of $\eta_n(\overline{\mathbf{S}}_{n,1},...,\overline{\mathbf{S}}_{n,K_I}) = \widehat{E}_{n,\max}$, which is obtained by solving problem (P3). Since the transmission arrangement of the symbol-blocks on all the $N$ sub-carriers is determined by the energy interleaving tensor $\mathbf{B}$, the utility function of the $n$-th sub-carrier can also be expressed as $\eta_n(\mathbf{B})$. Furthermore, the optimal constellation rotation angles $\mathbf{\Theta}$ are also functions of the energy interleaving tensor $\mathbf{B}$. Therefore, the original optimisation problem (P1) can be simplified as
\begin{align}
\text{(P4): }\max_{\mathbf{B}} \ \widehat{E}=\sum\limits_{n=1}^N\eta_n(\mathbf{B}),\ \ \text{s.t. }\ \ \eqref{P1-B},\eqref{P1-C},\eqref{P1-D}.\label{P5}
\end{align}
in which the energy interleaver tensor $\mathbf{B}$ is the only parameter needs to be optimised for maximising the total energy transferred by the transmitter.

Since $\mathbf{B}$ is a binary tensor having all its elements either 0 or 1, the optimisation problem (P4) is an integer programming problem. The exhaustive searching can be relied upon for the sake of obtaining the optimal solution.

\subsubsection{Exhaustive Searching}

A specific WIT user $u^I_k$ requests $N$ symbol-blocks on the $N$ sub-carriers in total. A single symbol-block of this WIT user is only allowed to be transmitted on a single sub-carrier. Therefore, we have $N!$ schemes for mapping the symbol-blocks of the WIT user $u^I_k$ to the sub-carriers. We denote $\mathbb{B}=\{\mathcal{B}_k\}$ having the size of $K_I\times 1$ as the feasible searching space for the energy interleaver tensor $\mathbf{B}$, where the element $\mathcal{B}_{k}\in\{1,\cdots,N!\}$ represents a symbol-block-to-sub-carrier mapping scheme for the WIT user $u^I_k$. We then traverse all the feasible mapping schemes for every WIT user and choose the optimal one for maximising the objective of (P4). The complexity of the exhaustive searching is as high as $\mathcal{O}((N!)^{K_I})$. Although $K_I$ may not be very large in a power-domain NOMA systems, the complexity will be substantially increased, if we have more sub-carriers.

\subsubsection{Greedy Algorithm}

In order to reduce the complexity of the exhaustive searching, another greedy algorithm is proposed by sacrificing some optimality of the solution. The main steps of the greedy algorithm is summarised as below:

\begin{itemize}
	\item \textit{Step 1}: Initialise the optimal utility function of every sub-carrier as $\eta^*_n=0$ for $n=1,\cdots,N$. A binary matrix $\mathbf{V}=\{v_{k,n}\}$ having the size of $K_I\times N$ is defined. Its element $v_{k,n}=1$ indicates that the $n$-th symbol-block of $u^I_k$ has already been scheduled on a specific sub-carrier but $v_{k,n}=0$ represents that the $n$-th symbol-block has not been scheduled on any sub-carrier. All the elements in $\mathbf{V}$ are initialised as $v_{k,n}=0$. Only the symbol-blocks having the indicators of $\{v_{k,n} = 0 | k =1,\cdots,K_I, n = 1,\cdots,N\}$ can be scheduled on the free sub-carriers. With the aid of $\mathbf{V}$, it can be thus guaranteed that each symbol-block requested by $u^I_k$ is scheduled on a single sub-carrier.
	\item \textit{Step 2}: When we start to schedule the symbol-blocks of $K_I$ WIT users on the $n$-th sub-carrier, the first $(n-1)$ sub-carriers have already accepted the transmissions of $K_I$ symbol-blocks for each. Accordingly, the WIT user $u^I_k$ only has $N-(n-1)$ symbol-blocks to be scheduled, whose binary indicators in the matrix $\mathbf{V}$ are all zeros. Hence, we have $(N-n+1)^{K_I}$ possible schemes in total for scheduling the rest of symbol-blocks of all the WIT users on the $n$-th sub-carrier. We denote all these scheduling schemes as a set $\mathbb{G}= \{\mathcal{G}_p | p = 1,\cdots, (N-n+1)^{K_I}\}$.
	\item \textit{Step 3}: We then traverse all the scheduling schemes  in the set $\mathbb{G}$ and find the optimal one for maximising the objective of (P4). When a specific scheduling scheme $\mathcal{G}_p \in \mathbb{G}$  is chosen, the symbol-blocks scheduled on the $n$-th sub-carrier are thus determined, which can be denoted as $\{\overline{\mathbf{S}}_{n,1},\cdots,\overline{\mathbf{S}}_{n,K_I}\}$. Therefore, we are capable of calculating the utility function $\eta_n(\overline{\mathbf{S}}_{n,1},\cdots,\overline{\mathbf{S}}_{n,K_I})$ of the $n$-th sub-carrier. Finally, the optimal scheduling scheme on the $n$-th sub-carrier can be found for the sake of maximising its utility function. In order to prevent these symbol-blocks from being scheduled again, we update the corresponding indicators in the matrix $\mathbf{V}$.
\end{itemize}

The pseudo code of the greedy algorithm is provided in Algorithm \ref{alg3}. The proposed greedy algorithm concludes after $N$ iterations. Its complexity is $\mathcal{O}(\sum\limits_{i=1}^Ni^{K_I})$, which is substantially lower than $\mathcal{O}((N!)^{K_I})$ of the exhaustive searching.

\begin{algorithm}[t]
\setstretch{1.0}
\caption{Greedy algorithm}
\label{alg3}
\small
\textbf{INPUT:}
User number $K_I$;
Sub-carrier number $N$;
Symbol blocks $\mathbf{S}_{1,1},\cdots,\mathbf{S}_{N,K_I}$;\\
\textbf{OUTPUT:}
Optimal energy interleaving tensor $\mathbf{B}^{*}$.
\begin{algorithmic}[1]
\STATE \textbf{Initialise} Optimal utility functions $\eta_n^{*}\leftarrow 0(1\le n\le N)$; Sub-carrier label $n\leftarrow 1$; Symbol block chosen matrix $\mathbf{V}=\{v_{k,n}\}\leftarrow \mathbf{0}$ and $\mathbf{V}\in \mathcal{Z}^{K_I\times N}$;
\WHILE {$n\le N$}
\STATE Initialise a set $\mathbb{G}$ consists of $(N-n+1)^{K_I}$ different combinations $\mathcal{G}_p (1\le p\le (N-n+1)^{K_I})$ for $K_I$ users, while each combination consists of $K_I$ different symbol-blocks $\mathbf{S}_{n_k,k}(1\le k\le K_I,\ 1\le n_k\le N)$ chosen from $K_I$ different users satisfying $v_{k,n_k}=0$; Combination chosen label $p\leftarrow 1$; Optimal combination $\mathcal{G}^{*}$;
\WHILE {$p\le (N-n+1)^{K_I}$}
\STATE Choose the combination $\mathcal{G}_{p}$;
\STATE Obtain $\mathbf{S}_{n_k,k}$ for each $u^I_k$ according to $\mathcal{G}_{p}$, let $\overline{\mathbf{S}}_{n,k}\leftarrow \mathbf{S}_{n_k,k}$;
\STATE Obtain constellation rotation angles $\{\theta_{n,k}(1\le k\le K_I)\}$ according to Algorithm \ref{alg1};
\STATE Obtain utility function $\eta_n(\overline{\mathbf{S}}_{n,1},\cdots,\overline{\mathbf{S}}_{n,K_I})$ by solving P3;
\STATE If $\eta_n(\overline{\mathbf{S}}_{n,1},\cdots,\overline{\mathbf{S}}_{n,K_I})>\eta_n^{*}$, then $\eta_n^{*}\leftarrow \eta_n(\overline{\mathbf{S}}_{n,1},\cdots,\overline{\mathbf{S}}_{n,K_I})$ and $\mathcal{G}^{*}\leftarrow \mathcal{G}_p$; $n_k^{*}\leftarrow n_k(1\le k\le K_I)$;
\STATE $p\leftarrow p+1$;
\ENDWHILE
\STATE $\{b^{*}_{k,m,n}\leftarrow 1 | m=n_k^{*}, k=1,\cdots,K_I\}$ and $\{b^{*}_{k,m,n}\leftarrow  0 |  m\neq n_k^{*}, k=1,\cdots,K_I\}$;
\STATE $v_{k,n_k^{*}}\leftarrow 1$; $n\leftarrow n+1$;
\ENDWHILE
\STATE Obtain the optimal energy interleaving tensor $\mathbf{B}^{*}$.
\end{algorithmic}
\end{algorithm}

\section{Simulation Results}

Without specific statement, the parameters in our simulation are set as follows: We have $K_I=3$ WIT users and $K_E=1$ WPT user in the NOMA-SWIPT system. The classic additive white Gaussian noise (AWGN) channels are conceived between the transmitter and the WIT/WPT users, while the noise power is set to be $\sigma^2=-80$ dBm. The channel power gain between the WPT user and the H-AP is $-30$ dB, while those between the WIT users and the H-AP are $\{-53,-60,-70\}$ dB, respectively. The transmitter has $10^6$ modulated symbols to be transmitted to every WIT user, which are randomly generated by the QAM based modulator. The symbol duration is set to be $T=10^{-6}$ s. There are 10 sub-carriers in the simulation. The total transmit power on every sub-carrier is all $P_n=1$ W.

\subsection{Convergence and Validity}

\begin{figure}[t]
  \centering
  \includegraphics[width=0.8\linewidth]{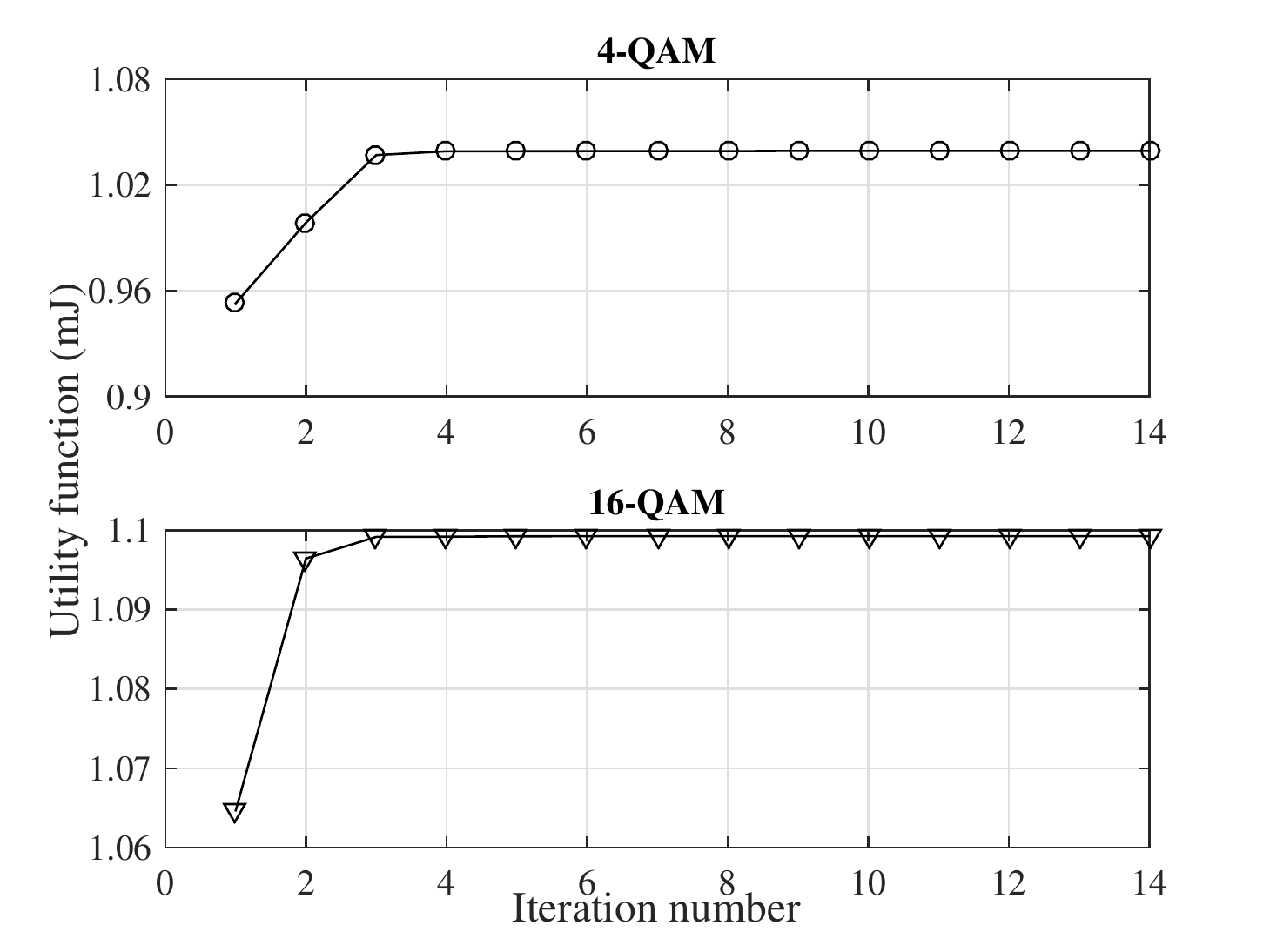}
  \setlength{\belowcaptionskip}{-12pt}
  \caption{Convergence of Algorithm.\ref{alg1}}\label{fig:1a}
\end{figure}
\begin{figure}[t]
  \centering
  \includegraphics[width=0.8\linewidth]{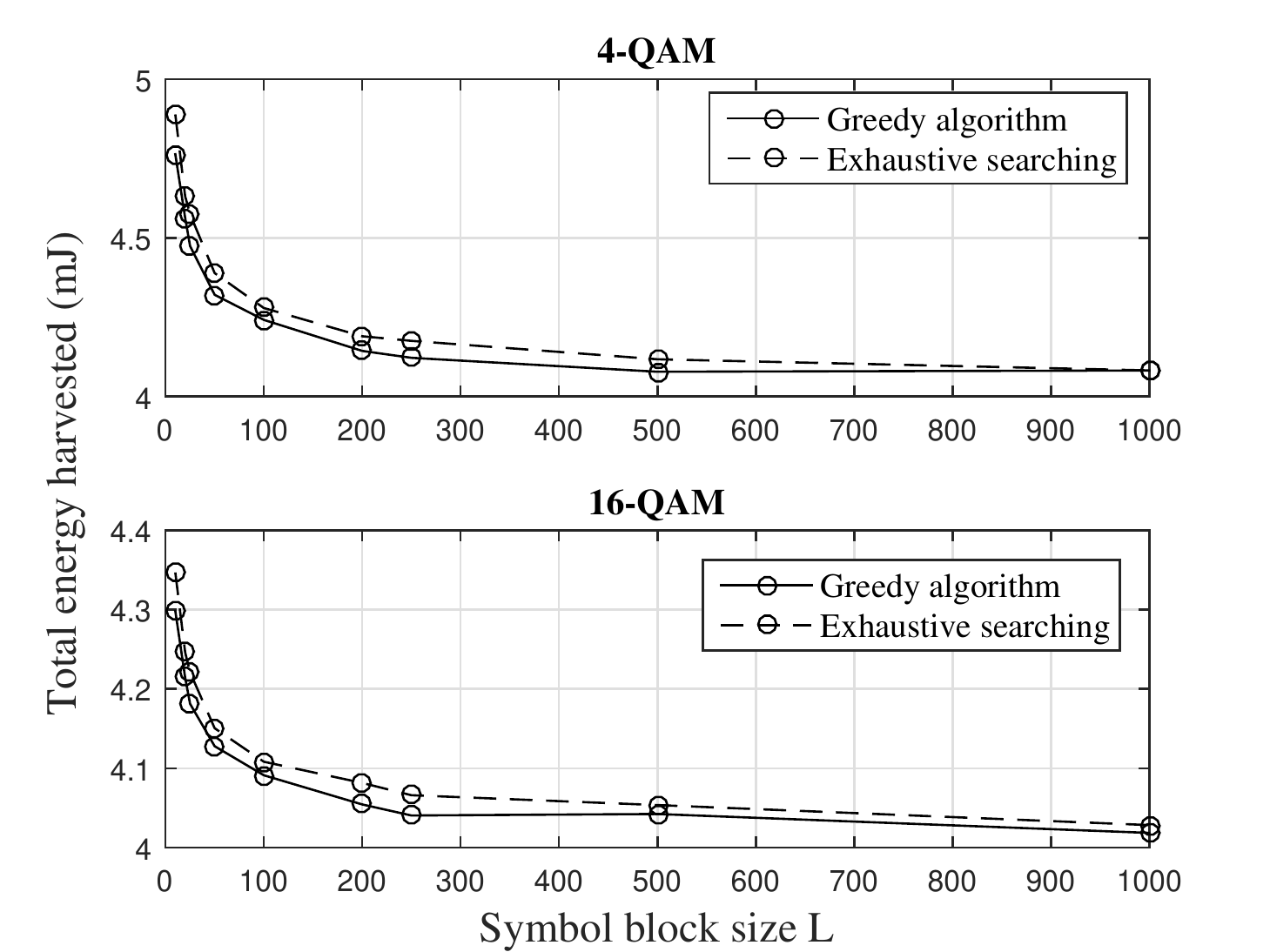}
  \setlength{\belowcaptionskip}{-12pt}
  \caption{Validity of Algorithm.\ref{alg3}}\label{fig:1b}
\end{figure}

We firstly investigate the convergence of Algorithm \ref{alg1} in Fig. \ref{fig:1a}, where we maximise the energy carried by a specific sub-carrier by solving the optimisation problem (P3). Observe from Fig. \ref{fig:1a} that Algorithm \ref{alg1} converges within 5 iterations at most, when the 4-QAM and 16-QAM are adopted.  Furthermore, we compare the  WPT performance of the greedy algorithm and the exhaustive searching based algorithm for solving the problem (P4) in Fig. \ref{fig:1b}, where we consider $N=4$ sub-carriers in total. Observe from Fig. \ref{fig:1b} that the WPT performance of the greedy algorithm is almost the same as that of the exhaustive searching. Therefore, in the rest of simulation, the greedy algorithm is adopted for solving the optimisation problem (P1), since its complexity is far lower than the exhaustive searching based counterpart.

\subsection{WPT Performance}

\begin{figure}[t]
  \centering
  \includegraphics[width=0.8\linewidth]{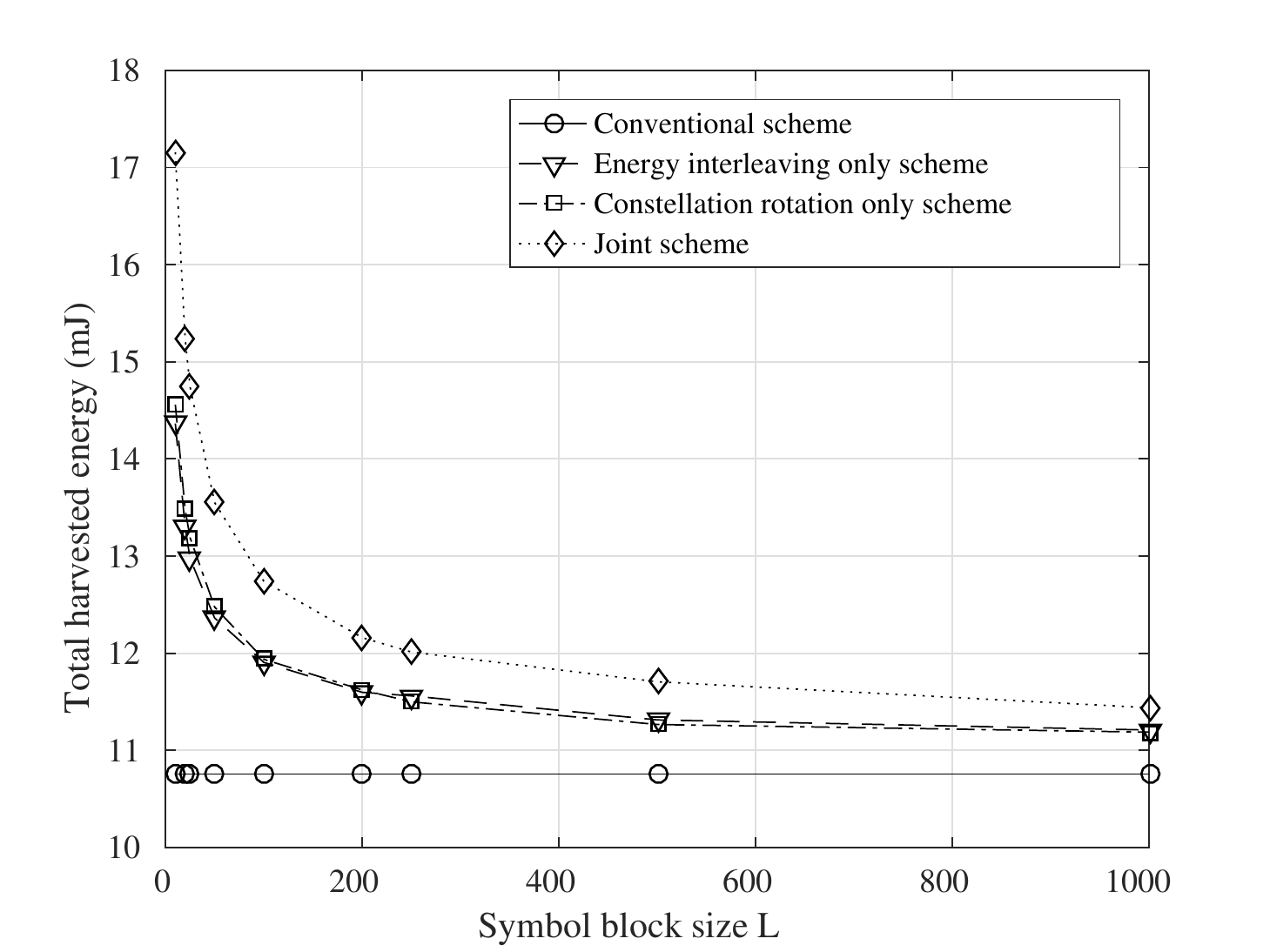}
  \setlength{\belowcaptionskip}{-12pt}
  \caption{WPT performance with 4-QAM modulation}\label{fig:3a}
\end{figure}
\begin{figure}[t]
  \centering
  \includegraphics[width=0.8\linewidth]{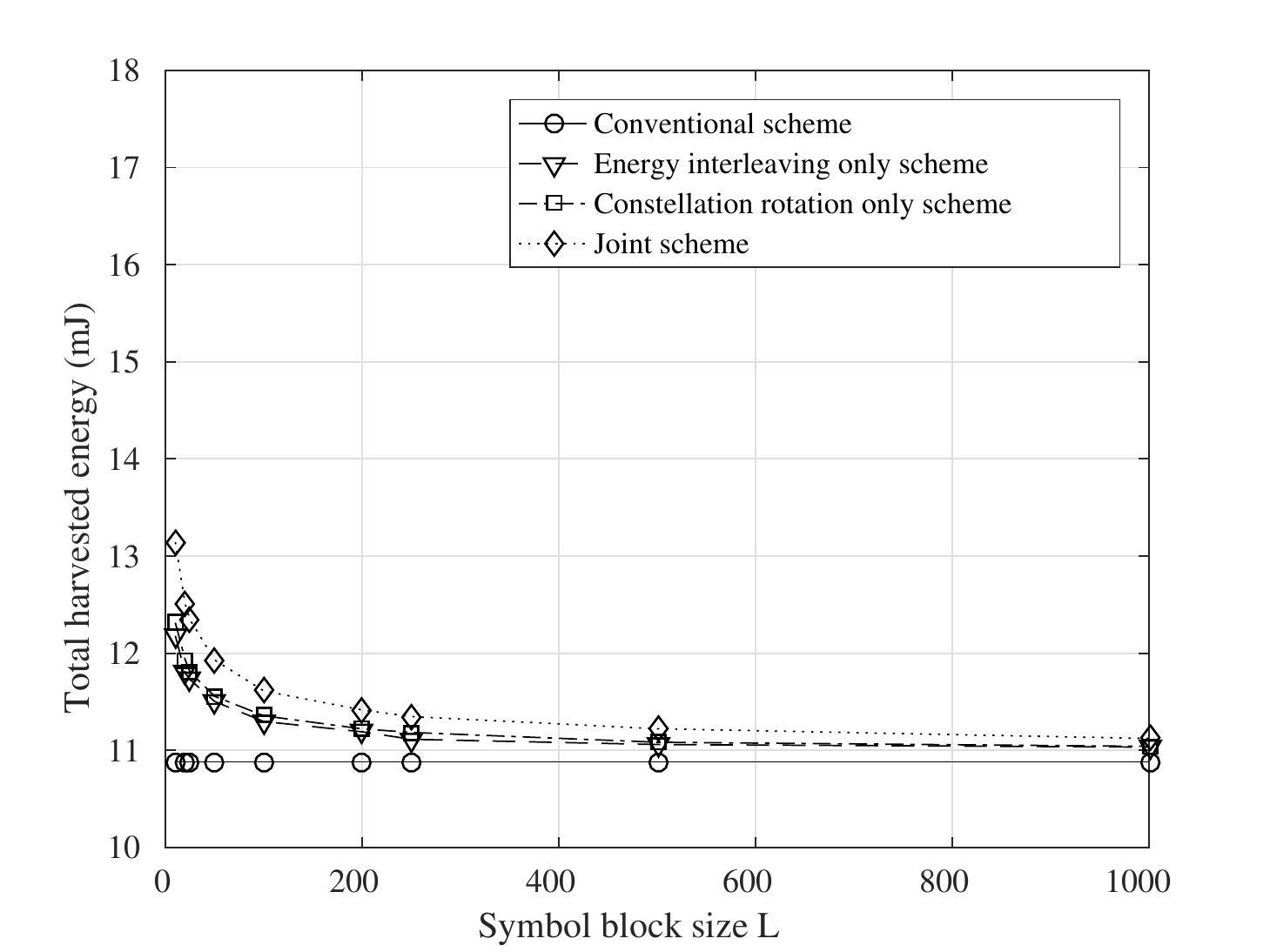}
  \setlength{\belowcaptionskip}{-12pt}
  \caption{WPT performance with 16-QAM modulation}\label{fig:3b}
\end{figure}

We evaluate the total energy  harvested by the WPT user in Fig. \ref{fig:3a}-\ref{fig:3b} by invoking the following different interleaving and modulation schemes:
\begin{itemize}
	\item \textit{Conventional scheme}: Neither energy interleaving nor constellation rotation is invoked;
	\item \textit{Energy interleaving only scheme}: constellation rotation is not invoked;
	\item \textit{Constellation rotation only scheme}: energy interleaving is not invoked;
	\item \textit{Joint scheme}: both interleaving and constellation rotation are invoked.
\end{itemize}
Observe from Fig. \ref{fig:3a}-\ref{fig:3b} that the \textit{joint scheme} achieves the highest WPT performance for both the 4-QAM and 16-QAM. Moreover, the \textit{energy interleaving only scheme} and the \textit{constellation rotation only scheme} achieve the moderate WPT performance, which is still much higher than the \textit{conventional scheme}. Furthermore, the WPT performance of the \textit{conventional scheme} keeps constant but that of all the other three schemes all reduces, as the symbol-block size $L$ increases. This is because a lower value of $L$ represents that the transmitter may update the energy interleaver and the constellation rotator more frequently, which results in more constructive symbol superposition. However, when we update the energy interleaver and the constellation rotator more frequently, more control signalling has to be transmitted to the WIT receivers, since the knowledge of the energy interleaver and the constellation rotators is indispensable for the demodulation and deinterleaving. Therefore, it's important to design an appropriate symbol-block size $L$ for the sake of compromising between the WPT performance and the control signalling overhead. By contrast, the WPT performance of the \textit{conventional scheme} is not affected by the symbol-block size $L$.

Moreover, observe from Fig. \ref{fig:3a}-\ref{fig:3b} that the 4-QAM based modulator achieves a higher WPT performance than the 16-QAM based modulators. According to the power allocation scheme of \eqref{Pkconstraint2}, when the higher order modulation is adopted, $A_{\max}$ also becomes higher, which indicates that the power difference among the symbols requested by different WIT users is enlarged. As a result, when we have a fixed total transmit power, the WPT performance gain incurred by the energy interleaving and the constellation rotation becomes lower, since the energy carried by a symbol having a very high power cannot be substantially degraded by a symbol having a very low power after their superposition. Under this consideration, if we want to achieve a higher energy harvesting gain, a lower rate is a better choice.

\begin{figure}[t]
\setlength{\belowcaptionskip}{-1cm}
  \setlength{\abovecaptionskip}{0.2cm}
\centering
  \includegraphics[width=0.8\linewidth]{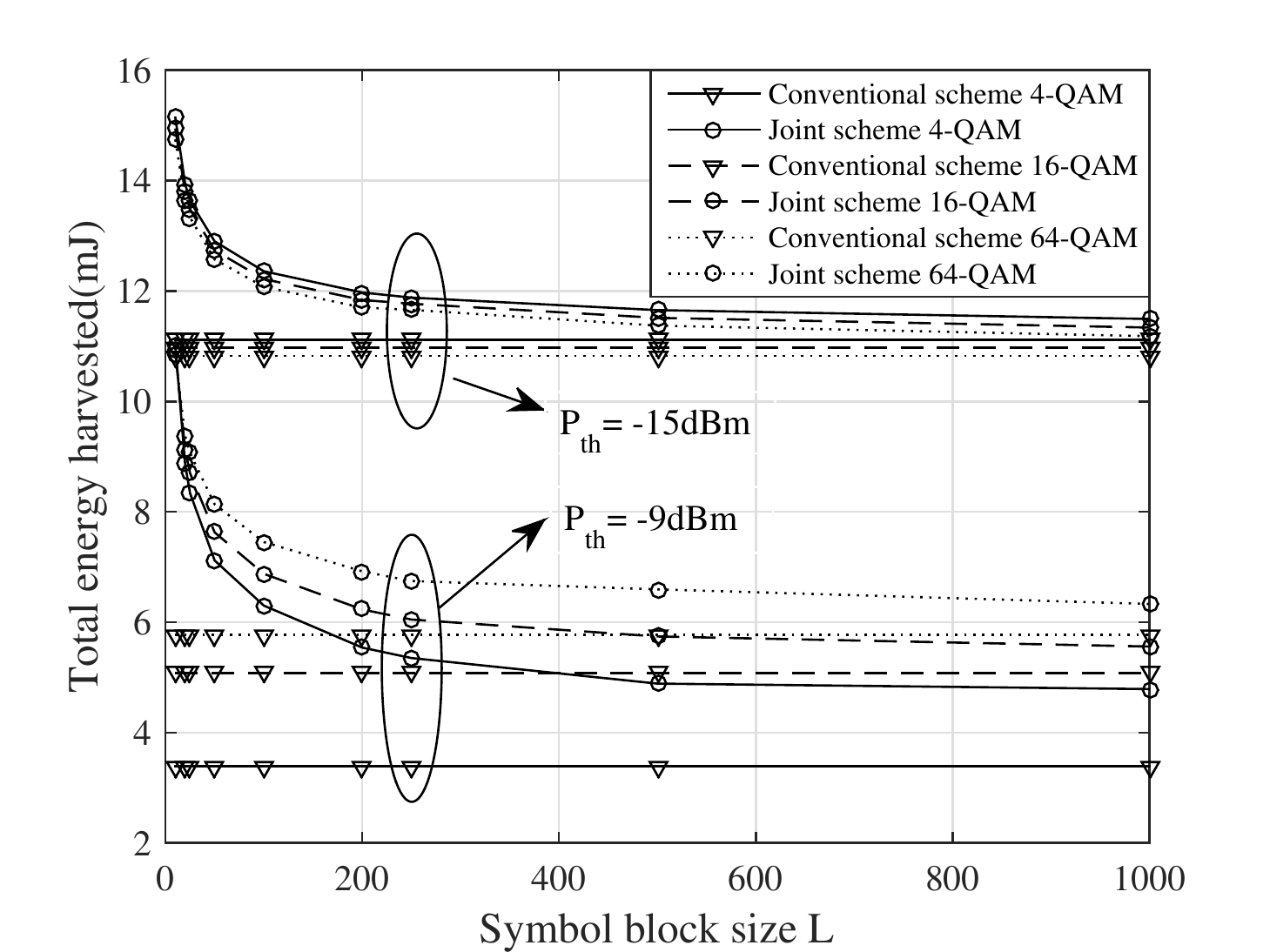}
  \caption{Comparasion among modulation schemes with different $P_\mathrm{th}$}\label{fig:7}
\end{figure}

\subsection{Sensitivity of Energy Harvester}
We investigate the impact of energy harvester's sensitivity $P_\mathrm{th}$ in Fig. \ref{fig:7}.  The power allocation scheme is derived by substituting $A_{\max}=\sqrt{98}$ of the 64-QAM into \eqref{Pkconstraint2}. The same power allocation scheme is also invoked for both the 4-QAM and 16-QAM in order to minimise the effect of the power difference on the attainable WPT performance. Observe from Fig. \ref{fig:7} that when the energy harvester's sensitivity is low, say $P_\mathrm{th} = -15$ dBm, the 4-QAM achieves the highest WPT performance, while the 64-QAM achieves the lowest WPT performance. When $P_\mathrm{th}$ increases to -9 dBm, we observe an inverse trend from Fig. \ref{fig:7}. This is because a modulation scheme of high order may generate a lot of symbols carrying lower power than the rectifier's sensitivity. The energy carried by these symbols cannot be harvested by the WPT user. By contrast, a modulation scheme of lower order generate symbols carrying almost identical power. If the sensitivity is low, all the symbols can be relied upon for energy harvesting. However, when the sensitivity is high, none of the symbols generated by a low-order modulation scheme are capable of delivering energy to the WPT users. By contrast, since a high-order modulation scheme may generate some symbols carrying very high power, these symbols can still deliver some energy to the WPT users.

\begin{figure}[t]
  \centering
  \includegraphics[width=0.8\linewidth]{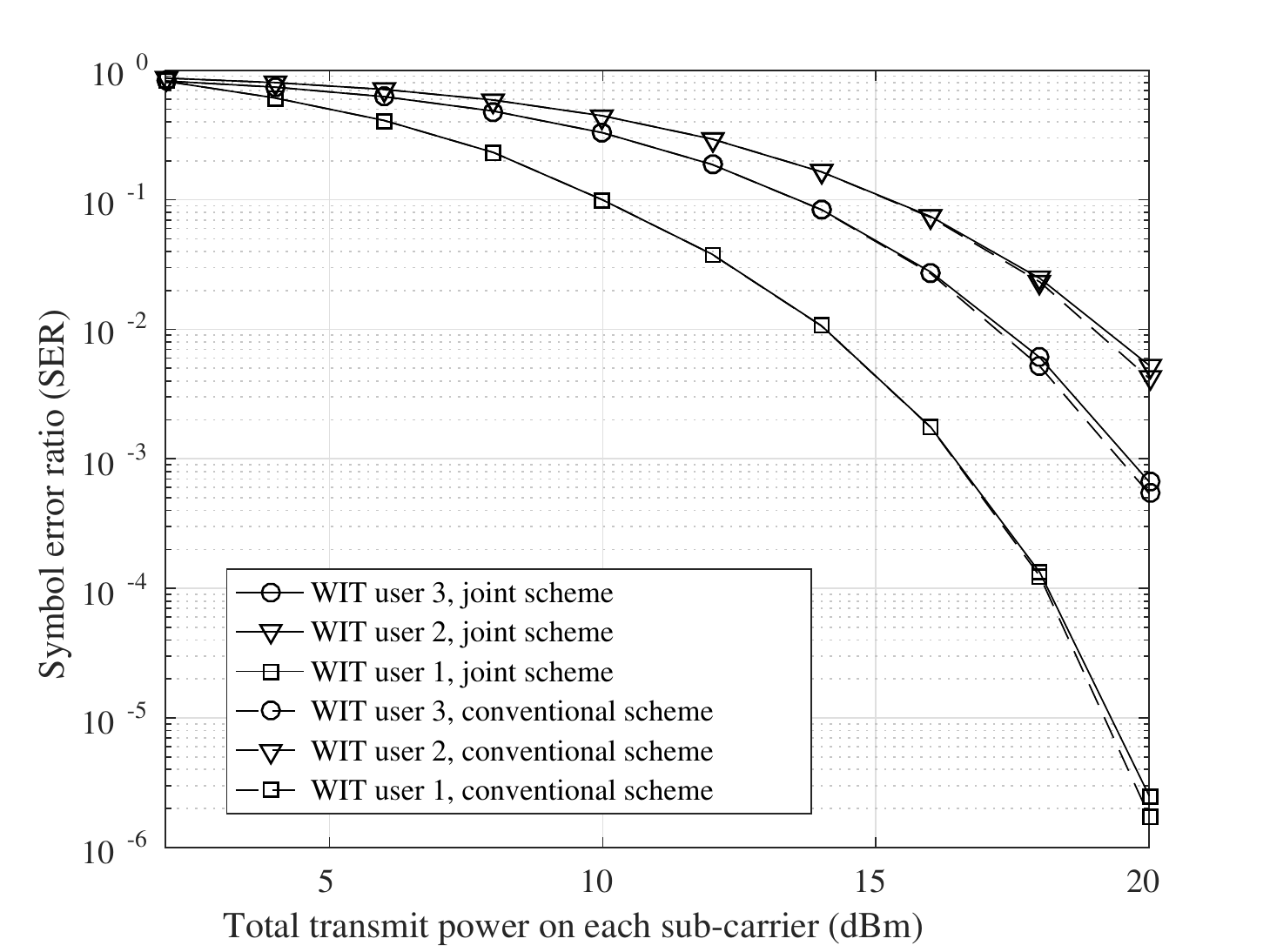}
  \setlength{\belowcaptionskip}{-12pt}
  \caption{WIT performance with 4-QAM modulation}\label{fig:9a}
\end{figure}
\begin{figure}[t]
  \centering
  \includegraphics[width=0.8\linewidth]{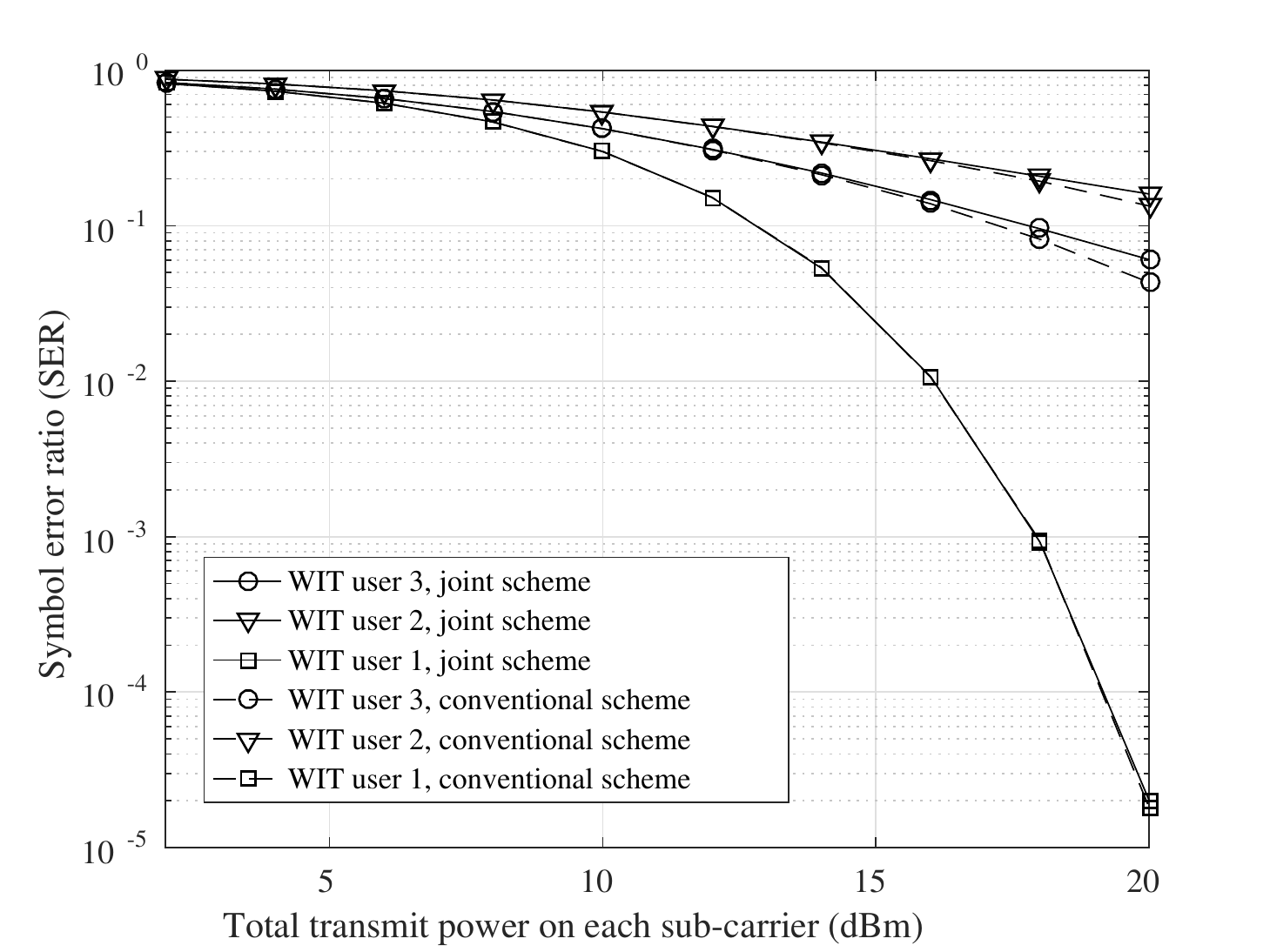}
  \setlength{\belowcaptionskip}{-12pt}
  \caption{WIT performance with 16-QAM modulation}\label{fig:9b}
\end{figure}

\begin{figure*}[t]
	\subfigure[16-QAM]{\includegraphics[width=0.32\linewidth]{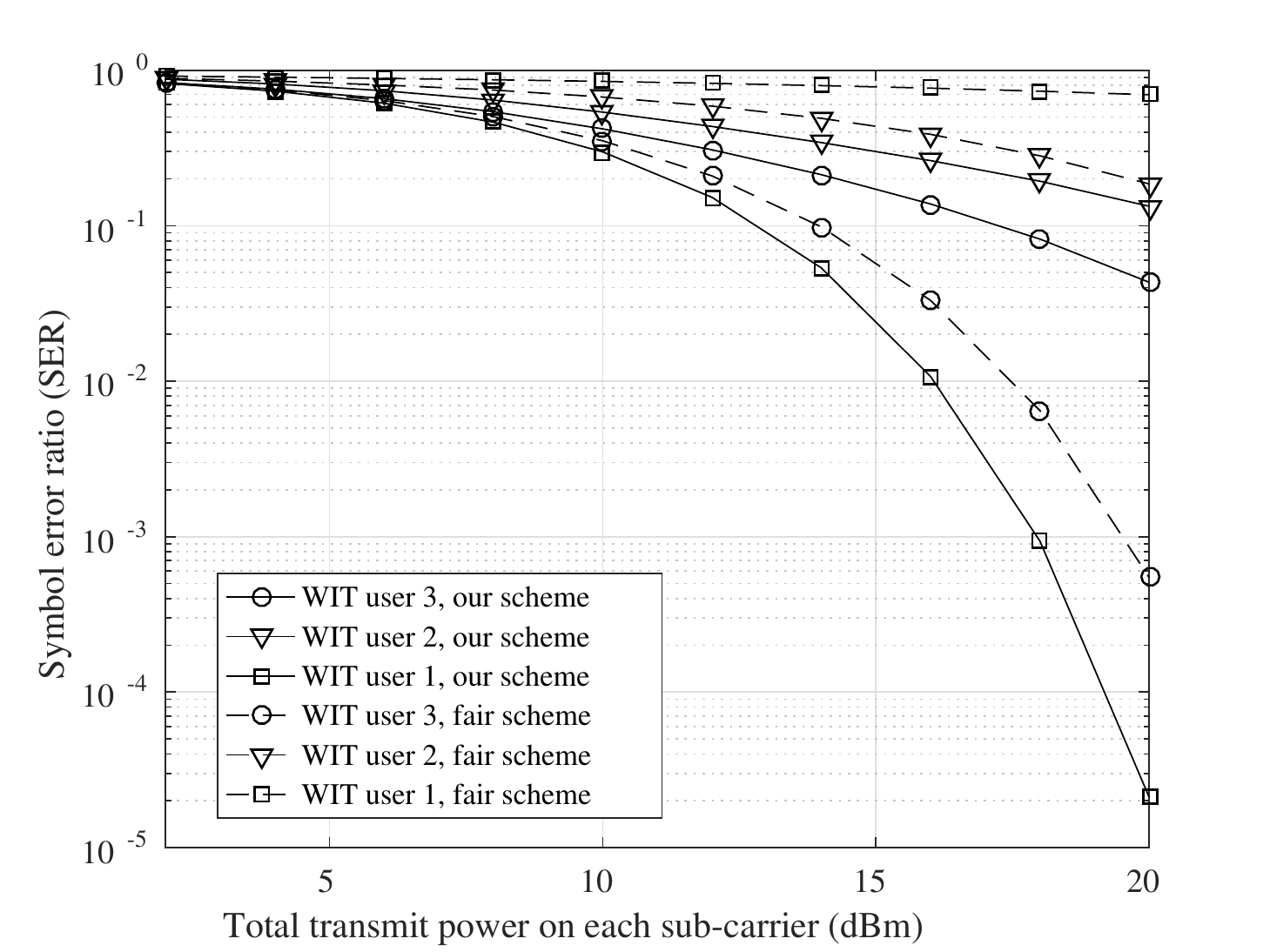}}
	\hfill
	\subfigure[64-QAM]{\includegraphics[width=0.32\linewidth]{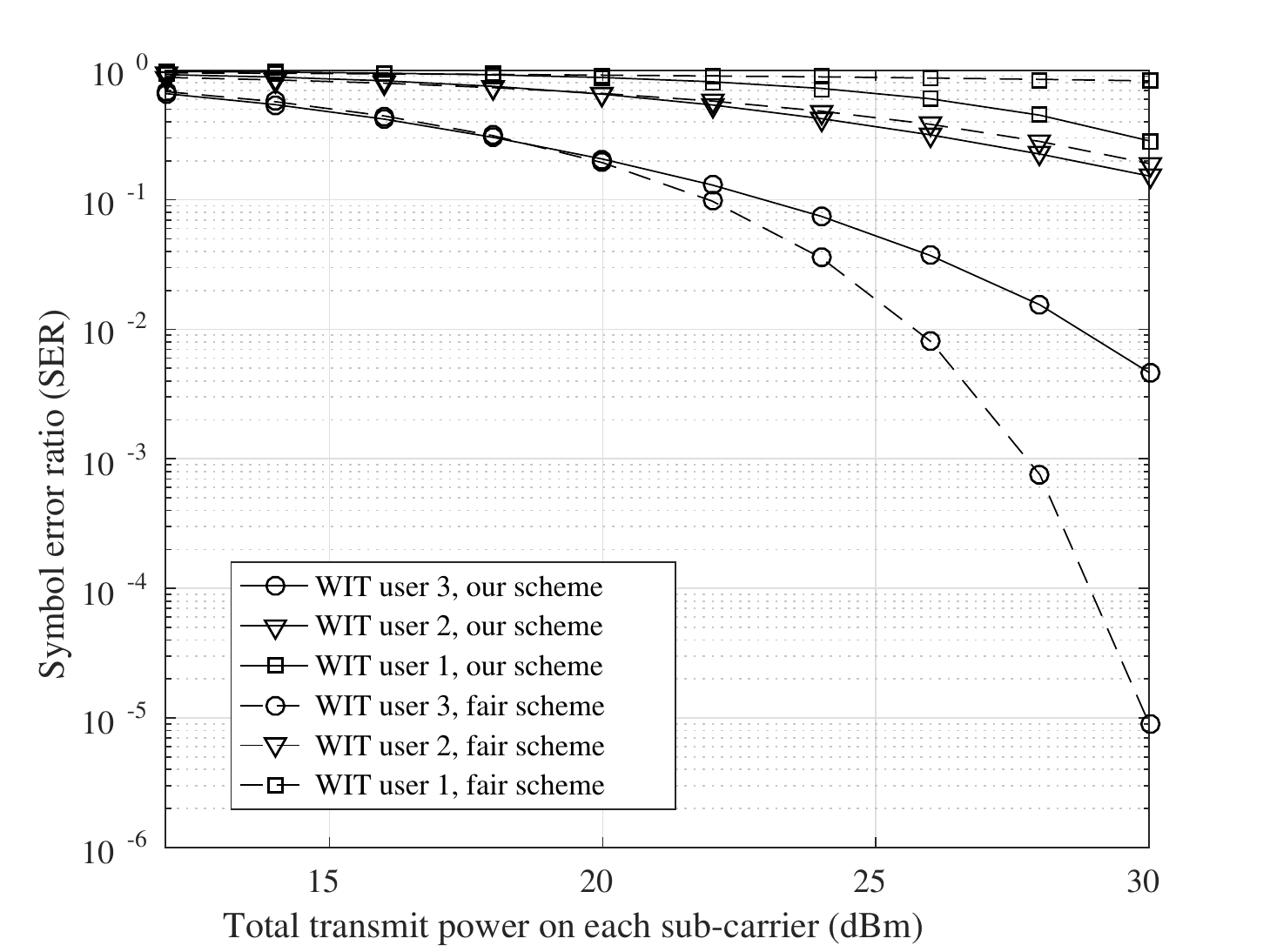}}
    \hfill
	\subfigure[Average SER among WIT users]{\includegraphics[width=0.32\linewidth]{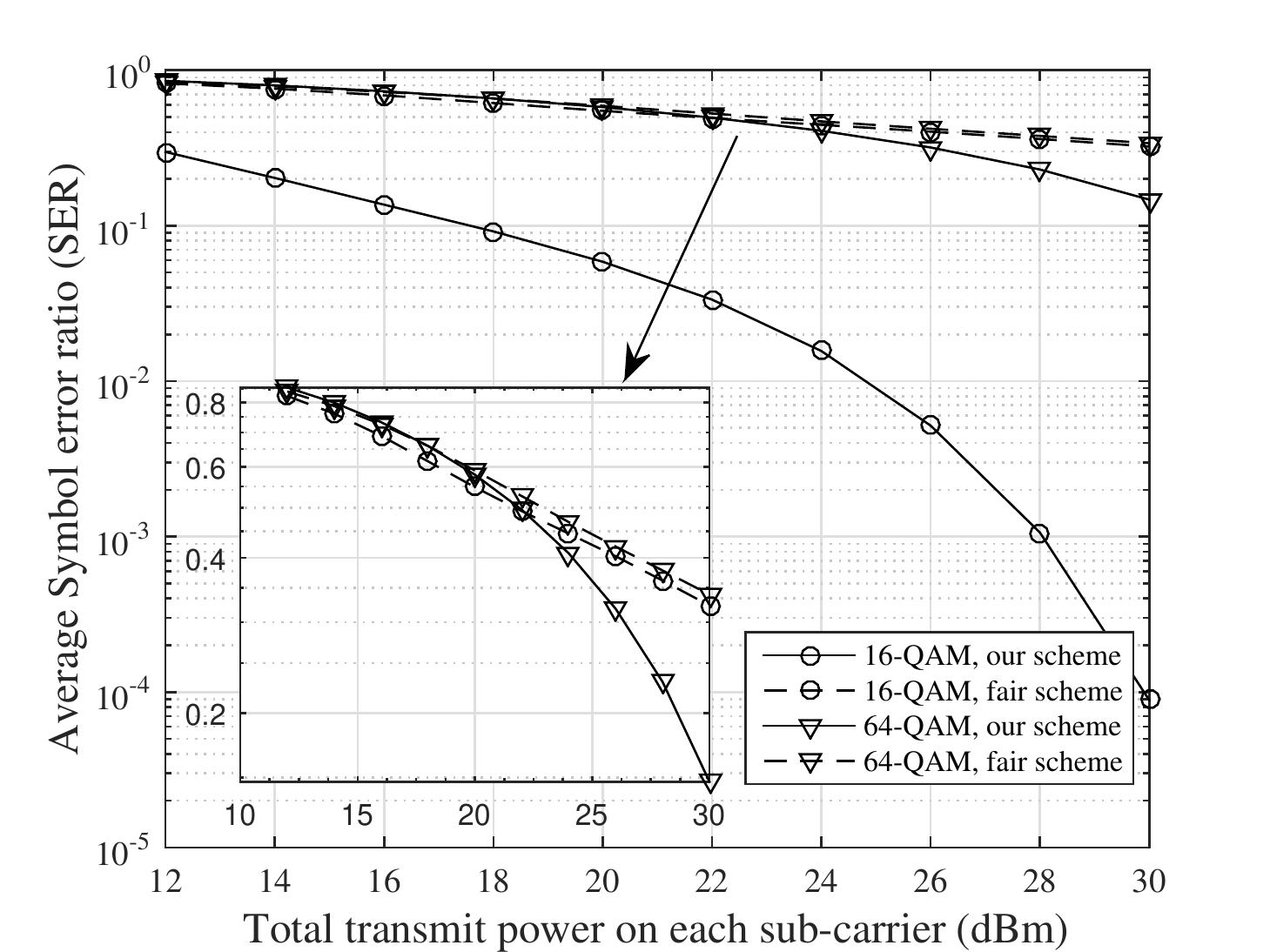}}
	\caption{WIT performance with different power allocation schemes}\label{fig:10}
\end{figure*}

\subsection{WIT Performance}

Finally, we plot the SER performance of the joint and conventional schemes versus the total transmit power on each sub-carrier in Fig. \ref{fig:9a}-\ref{fig:9b}, when 4-QAM and 16-QAM are adopted, respectively. In this simulation, the transmitter has $10^8$ modulated symbols to be transmitted to each WIT user in order to achieve more accurate results. Observe from Fig. \ref{fig:9a}-\ref{fig:9b} that the joint scheme has almost the same SER performance as the conventional scheme, when the 4-QAM is adopted. When the 16-QAM are adopted, the joint scheme only suffers from a tiny degradation of the SER performance, when compared to the conventional counterpart.

We also compared the SER performance in Fig. \ref{fig:10} between our scheme and another fair scheme proposed  in \cite{5959}, where the power allocation scheme is designed by letting the signal to interference and noise ratio (SINR) of all the WIT users be the same. Observe from Fig. \ref{fig:10} that when 16-QAM is adopted, for WIT user 3 having the worst channel condition, our scheme achieves a higher SER than the counterpart. However, for WIT user 1 having the best channel condition, our power allocation scheme achieves a much lower SER than the fair scheme.  When 64-QAM is adopted, our scheme achieves lower SER differences among WIT users than the counterpart. Moreover, observe from Fig. \ref{fig:10}(c) that our power allocation scheme achieves lower average SER among all the WIT users than the fair scheme.  In a nutshell, our power allocation scheme outperforms the benchmark of \cite{5959} in terms of the SER fairness.

\subsection{Spectrum Efficiency}
Let us now discuss the spectrum efficiency of our NOMA-SWIPT system. As for the WIT user $u^I_k$, the total achievable throughput is formulated as
\begin{align}
R_k=\sum\limits_{n=1}^{N}B\log(1+\frac{h_{n,k}P_{n,k}}{\sum_{i=1}^{k-1}h_{n,i}P_{n,i}+\sigma^2})\notag
\end{align}
where $B$ is the bandwidth of a single sub-carrier. Therefore, the spectrum efficiency of the system is derived as
\begin{align}
\kappa=\frac{\sum_{k=1}^{K_I}R_k}{NB}\notag
\end{align}
In our simulation, the spectrum efficiency is thus calculated as $7.47 \text{ bps/Hz}$, which is around  $17\%$ higher than  $6.36 \text{ bps/Hz}$  of the OFDMA based counterpart. Meanwhile, observe from Fig. \ref{fig:3a}-\ref{fig:3b} that if the symbol block size is set to be $L=100$, our joint design is capable of achieving $18\%$ and $6\%$ higher  WPT performance than the conventional scheme, when 4-QAM and 16-QAM are adopted, respectively.


\section{Conclusion}

In this paper, we jointly design the energy interleaving and the constellation rotation in a multi-user NOMA-SWIPT system. In our design, the energy interleaving tensor as well as the constellation rotation angles are optimised for maximising the actual energy carried by the superposition signals. Furthermore, a power allocation scheme among the symbols requested by the WIT users is proposed for the sake of reducing the SER of the WIT users, when the SIC is adopted for the demodulation of the superposition symbols. The simulation results demonstrate that our joint design is capable of substantially increasing the attainable WPT performance, while the SER degradation can be rigorously controlled. Furthermore, our simulation results also demonstrate that for a more sensitive energy harvester associated with a higher activation threshold, the higher-order modulation scheme outperforms its low-order counterparts. By contrast, for an energy harvester having a lower activation threshold, the lower-order modulation scheme performs better than its high-order counterparts.

\appendices
 \section{}
For an arbitrary WIT user $u^I_k$, the distortion constraint \eqref{eq:7} has to be satisfied in order to reduce the distortion incurred by the other $(k-1)$ symbols requested by the WIT users $\{u^I_1,\cdots,u^I_{k-1}\}$ in the SIC aided demodulation. In the worst case, the symbol distortion $\widehat{\xi}_{n,k,l}$ of \eqref{xi} is maximised, when its in-phase component $\widehat{\xi}_{n,k,l}^I$ and quadrature component $\widehat{\xi}_{n,k,l}^Q$ both achieve their maximums. As a result, the symbols transmitted to the WIT users $\{u_1^I,\cdots, u_{k-1}^I\}$ all have the highest normalised amplitudes, say $\widehat{A}_{n,1,l}=\cdots=\widehat{A}_{n,k-1,l} = A_{\max}$, while they also have the same phase, say $\widehat{\phi}_{n,1,l} =\cdots= \widehat{\phi}_{n,k-1,l}$. Since we have
\begin{align}
&\max\ \cos(\widehat{\phi}_{n,i,l}+\theta_{n,i}-\theta_{n,k})=1,\notag\\
&\max\ \sin(\widehat{\phi}_{n,i,l}+\theta_{n,i}-\theta_{n,k})=1,
\end{align}
the symbol distortion constraint \eqref{eq:7} can be reformulated as
\begin{align}
d_{n,k}\ge A_{\max}\sum\limits_{i=1}^{k-1}d_{n,i},\ \  k=2,\cdots,K_I.
\end{align}
Since $d_{n,k}=\sqrt{\frac{3P_{n,k}}{M-1}}$ is a linear function of $\sqrt{P_{n,k}}$, we can readily obtain the constraint \eqref{eq:12} on the transmit power of the symbols requested by different WIT users.
\bibliography{Modulation-NOMA}

\end{document}